%% file: main.tex
\documentclass[twocolumn,epjc3]{svjour3}          

\RequirePackage[T1]{fontenc}

\smartqed  

\input{packages}
\usepackage{ATLAS_macros}
\RequirePackage{flushend}
\RequirePackage[numbers,sort&compress]{natbib}

\journalname{Eur. Phys. J. C}

\begin{document}

\title{Seeking a coherent explanation of LHC excesses for compressed spectra}

\author{
        Diyar Agin\,\orcid{0009-0007-7146-5242}\thanksref{e1,addr}
        \and
        Benjamin Fuks\,\orcid{0000-0002-0041-0566}\thanksref{e2,addr}
        \and
        Mark D. Goodsell\,\orcid{0000-0002-6000-9467}\thanksref{e3,addr}
        \and
        Taylor Murphy\,\orcid{0000-0002-3215-9652}\thanksref{e4,addr}
}

\thankstext{e1}{\href{mailto:dagin@lpthe.jussieu.fr}{dagin@lpthe.jussieu.fr}}
\thankstext{e2}{\href{mailto:fuks@lpthe.jussieu.fr}{fuks@lpthe.jussieu.fr}}
\thankstext{e3}{\href{mailto:goodsell@lpthe.jussieu.fr}{goodsell@lpthe.jussieu.fr}}
\thankstext{e4}{\href{mailto:murphy@lpthe.jussieu.fr}{murphy@lpthe.jussieu.fr}}

\institute{Laboratoire de Physique Th\'{e}orique et Hautes \'{E}nergies (LPTHE),\\ UMR 7589, Sorbonne Universit\'{e} \& CNRS,\\ 4 place Jussieu, 75252 Paris Cedex 05, France \label{addr}}

\date{Accepted 6 November 2024}

\maketitle

\input{Sections/abstract} 

\input{Sections/1_Intro}
\input{Sections/2_Recasts}
\input{Sections/3_MSSM}
\input{Sections/4_NMSSM}
\input{Sections/5_Non_SUSY}
\input{Sections/6_Conclusion}

\begin{acknowledgements}

B. F., M. D. G., and T. M. are supported in part by Grant ANR-21-CE31-0013, Project DMwithLLPatLHC, from the \emph{Agence Nationale de la Recherche} (ANR), France. We thank Jack Y. Araz for correspondence about \spey.

\end{acknowledgements}

\appendix

\input{Sections/A_ATLAS_3l_extra_tables}

\bibliographystyle{JHEP}
\bibliography{Bibliography/bibliography}

\end{document}

%% file: packages.tex
\usepackage[dvipsnames]{xcolor} 

\definecolor{mpl-orange}{HTML}{FFA500}
\definecolor{mpl-green}{HTML}{008000}

\usepackage{afterpage}
\usepackage{amsfonts, amssymb}
\usepackage{anyfontsize}
\usepackage{booktabs} 
\usepackage[colorlinks,citecolor=blue,urlcolor=blue,linkcolor=blue]{hyperref} 
\usepackage{makecell}
\usepackage{mathtools}
\usepackage{multirow}
\usepackage{physics}
\usepackage{slashed}
\usepackage{xspace} 
\usepackage[english]{babel}
\usepackage[section]{placeins}

\usepackage{newtxtext,newtxmath}

\newcommand{\ie}{\textit{i.e.}}
\newcommand{\eg}{\textit{e.g.}}
\def\frules{{\sc FeynRules}\xspace}
\def\MadGraph{{\sc MG5\_aMC}\xspace}
\newcommand{\madgraph}{{\sc MG5\_aMC}\xspace}

\newcommand{\madanalysis}{{\sc Mad\-A\-na\-ly\-sis\,5}\xspace}
\newcommand{\pythia}{{\sc Pythia\,8}\xspace}
\newcommand{\spey}{{\sc Spey}\xspace}
\newcommand{\pyhf}{{\sc PyHF}\xspace}
\newcommand{\hackanalysis}{{\sc HackAnalysis}\xspace}
\newcommand{\BSMArt}{{\sc BSMArt}\xspace}
\newcommand{\YODA}{{\sc YODA}\xspace}

\newcommand{\MET}{$E_{\text{T}}^{\text{miss}}$\ }

\def\d{{\mathrm{d}}} 
\def\ii{{\text{i}}}

\makeatletter
  \newcommand*{\transpose}{{\mathpalette\@transpose{}}}
  \newcommand*{\@transpose}[2]{\raisebox{\depth}{$\m@th#1\intercal$}}
\makeatother

\newcommand{\orcid}[1]{\begingroup\href{https://orcid.org/#1}{\includegraphics[width=9pt]{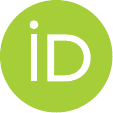}} \endgroup}
\newcommand{\tv}[1]{\overset{{}_{\,\scalebox{0.55}{$\shortrightarrow$}}}{#1}}
\newcommand{\overbar}[1]{\mkern 2mu\overline{\mkern-4mu#1\mkern-4mu}\mkern 2mu}
\newcommand{\shortrightarrow}{\mathrel{\scalebox{0.75}{$\rightarrow$}}}

\makeatletter
\renewcommand{\fnum@figure}{\textsc{\figurename~\thefigure}} 
\makeatother

\makeatletter
\renewcommand*\l@subsubsection[2]{}
\makeatother

%% file: Sections/abstract.tex
\begin{abstract}

The most recent searches by the ATLAS and CMS Collaborations in final states with soft leptons and missing transverse energy show mild excesses predominantly associated with dilepton invariant masses of about $10\text{--}20$~GeV, which can result from decays of electroweakinos that are heavier than the lightest neutralino by $\mathcal{O}(10)$~GeV. On the other hand, these analyses are insensitive to electroweakino mass splittings smaller than about 5~GeV. In previous work, we demonstrated that while recent searches in the monojet channel can exclude some of the smallest $\mathcal{O}(1)$~GeV mass splitting configurations for electroweakinos, they also exhibit excesses that can overlap with the soft-lepton excesses in certain models, including a simplified scenario with pure higgsinos. In this work we dive deeper into these excesses, studying the analyses in detail and exploring an array of models that go beyond the simplified scenarios considered by the experimental collaborations. We show that, in the Minimal Supersymmetric Standard Model, the overlapping excesses are not unique to the pure-higgsino limit, instead persisting in realistic parameter space featuring a bino-like lightest supersymmetric particle with some wino admixture. On the other hand, for the Next-to-Minimal Supersymmetric Standard Model with a singlino-like lightest supersymmetric particle and higgsino-like next-to-lightest supersymmetric particle(s), the excess in the two-lepton channel fits rather well with the parameter space predicting the correct relic abundance through freeze out, but the monojet fit is much poorer. Interestingly, the excesses either do not overlap or do not exist at all for two non-supersymmetric models seemingly capable of producing the correct final states.

\keywords{Physics beyond the Standard Model \and Soft leptons \and Monojets \and Supersymmetry}

\end{abstract}

%% file: Sections/1_Intro.tex
\section{Introduction}
\label{s1}

As experimental limits on classic supersymmetric scenarios grow ever tighter, analyses at the Large Hadron Collider (LHC) have increasingly targeted alternative scenarios that are more experimentally challenging, including those with light neutralinos and charginos (hereafter referred to collectively as electroweakinos) whose masses differ by $\mathcal{O}(10)~\text{GeV}$ or less (forming a so-called \emph{compressed spectrum}). Surprisingly, among the many analyses performed during Run 2 of the LHC, both the ATLAS \cite{ATLAS:2019lng,ATLAS:2021moa} and CMS \cite{CMS:2021edw} Collaborations have reported excesses in equivalent searches involving two or more soft leptons, interpreted in terms of simplified supersymmetric scenarios in which the light electroweakinos are composed mostly of higgsinos or the bino and wino. These interpretations are inspired by certain limits of the Minimal Supersymmetric Standard Model (MSSM). The excesses mostly appear in events containing two leptons ($2\ell$) with a dilepton invariant mass in the range 10--20~GeV, but very slight excesses are also visible in the ATLAS trilepton ($3\ell$) channel. It is moreover at least qualitatively clear, modulo differences in the model interpretations, that the ATLAS and CMS excesses appear in nearly the same regions of simplified higgsino and/or bino-wino MSSM parameter space; this overlap suggests that a model producing a good fit to one soft-lepton analysis may hope to do the same for the other search. These excesses have thus sparked the interest of the phenomenological community, with several papers exploring some consequences for those scenarios in combination with other observables \cite{Canepa:2020ntc,Buanes:2022wgm,Cao:2022htd,Domingo:2022pde,Barman:2022jdg,Baum:2023inl,Stark:2023ont,Baer:2023olq,Cao:2023juc,Ashanujjaman:2023tlj,Carpenter:2023agq,Baer:2023ech,Roy:2024yoh,Baer:2024kms,Chakraborti:2024pdn,Martin:2024pxx}.

In a previous work \cite{Agin:2023yoq}, we demonstrated that these dilepton excesses are surprisingly compatible with excesses in \emph{monojet} events, again appearing in analyses by both the ATLAS \cite{ATLAS:2021kxv} and CMS \cite{CMS:2021far} Collaborations. This was achieved by interpreting the monojet signals in terms of the simplified scenario used in both soft-lepton searches in which the lightest supersymmetric particle (LSP) $\tilde{\chi}^0_1$, the next-to-lightest supersymmetric particle (NLSP) $\tilde{\chi}^{\pm}_1$, and the second-lightest neutralino $\tilde{\chi}^0_2$ constitute a triplet of ``pure'' higgsinos (we refer to this as the ``simplified higgsino'' scenario). These findings raised several questions; in particular:
\begin{enumerate}
    \item Since the simplified higgsino scenario is really a toy model, what happens when we consider \emph{realistic} supersymmetric scenarios, within the MSSM or otherwise?
    \item Are there non-supersymmetric models that can explain the excesses?
    \item Can any of these models produce a dark matter candidate with suitable relic abundance?
\end{enumerate} 
In this work, we begin to address these three questions.

In the ATLAS and CMS compressed electroweakino analyses, the simplified models invariably produce dilepton events through the off-shell decay of a heavy neutralino to the LSP, $\tilde{\chi}_2^0 \to Z^* +\tilde{\chi}_1^0,\ Z^* \to \ell^+ \ell^-$. But it is also possible to produce dilepton events through decays of off-shell $W$ bosons through processes such as
\begin{align}
\nonumber    p p \to \tilde{\chi}_1^+ &\tilde{\chi}_1^-,\\ &\tilde{\chi}_1^\pm \to \tilde{\chi}_1^0 + W^{\pm,*},\ W^{\pm,*} \to \ell^\pm \nu.
\end{align}
In the simplified higgsino model these contribute a small but non-negligible amount to the signal. On the other hand, in the other scenario considered by the experimental collaborations, in which the LSP is predominantly bino and the NLSP is mostly wino (we refer to this as the ``simplified bino-wino'' scenario, though as we discuss below, it encompasses two cases), the cross section of the $pp\to \tilde{\chi}^+_1 \tilde{\chi}^-_1$ process is small enough that the experimental collaborations neglect its contribution in their simulations. Other models with the same pattern of decays could, however, produce such decays with larger rates and these processes could, in principle, provide an explanation of the observed excesses. Furthermore, \emph{two-body} decays could also provide soft dileptons, for example in models with new states that couple directly to leptons. Thus our investigation is aimed, in part, at understanding how the topology of a given $2\ell$ signal process affects its ability to fit the excesses observed by the ATLAS and CMS Collaborations. But this has to be done with due regard for the monojet analyses, which veto practically all visible leptons and therefore require any candidate model to produce monojet events through entirely different processes.

Meanwhile, it is generally accepted that models with a higgsino LSP much lighter than 0.5--1.0~TeV predict underabundant dark matter (assuming a standard cosmological history)~\cite{Baer:2011ab,Chakraborti:2017dpu}, which by itself motivates non-higgsino interpretations of the compressed LHC excesses. Clearly spurred (at least in part) by this fact, ATLAS recently performed a study in the phenomenological MSSM (pMSSM) examining limits on their own bino-wino scenarios, comparing the limits from soft-lepton searches to constraints from dark matter searches and elsewhere \cite{ATLAS:2024fyl}. But this analysis includes neither their own monojet search\footnote{We speculate that this omission arises because this analysis was performed by the exotica group, rather than by the supersymmetry group.} nor the CMS analog, in which the excess appears stronger. Thus the results are presented as limits, even though the mild soft-lepton excess can be discerned from their plots.

The starting point for this work is therefore a holistic (re)examination of the bino-wino MSSM scenario. We perform a scan over realistic MSSM parameter space and analyze the resulting points using recasts of all four relevant LHC analyses. Our recasts of the ATLAS $2\ell$ and $3\ell$ analyses are released alongside this work and documented and validated here for the first time. Further motivated by the dark matter problem, we also identify parameter space within the Next-to-Minimal Supersymmetric Standard Model (NMSSM) that may be suitable for the LHC excesses and perform a similar analysis. We then consider two non-supersymmetric beyond-the-Standard Model (BSM) scenarios that seem like good candidates and interestingly seek to explain the LHC excesses through processes markedly different from their supersymmetric competitors.

This paper is organized as follows. Section \ref{s2} describes the soft-lepton (and, briefly, monojet) analyses considered in this work, with particular detail devoted to the various excesses and the validation of our recasts. In Section \ref{s3} we begin to interpret the analyses through the lens of BSM physics, starting with a realistic MSSM scenario featuring a bino-wino LSP and decoupled higgsinos. We emphasize the differences between our scenario and the simplified bino-wino model considered by ATLAS, and explore the overlap between the multiple excesses and the parameter space compatible with the observed dark matter relic abundance. We consider the NMSSM in Section \ref{s4}, demonstrating that light higgsinos and an even lighter singlino can generate a $2\ell$ excess compatible with the dark matter relic abundance, but fails to predict enough monojet events. Finally, in Section \ref{s5}, we show that two seemingly well motivated non-supersymmetric models show either non-overlapping or nonexistent excesses, primarily due to their failure to reproduce the lepton kinematics of the supersymmetric models. Thus the interpretation of these analyses is highly sensitive to the topologies of the signal processes that generate soft lepton pairs. We summarize our findings and suggest future avenues of study in Section \ref{s6}.

%% file: Sections/2_Recasts.tex
\section{Discussion of experimental analyses}
\label{s2}

We begin by reviewing the experimental analyses of interest in this work. We detail the implementation of these analyses in \hackanalysis version 2~\cite{Goodsell:2024aig},\footnote{Available at \url{https://goodsell.pages.in2p3.fr/hackanalysis}.} which we use as part of a simulation and analysis toolchain to evaluate the compatibility of multiple BSM scenarios with the compressed LHC excesses.

\subsection{ATLAS-SUSY-2018-16: Soft dilepton + missing energy}
\label{s2.1}

\begin{figure}
    \centering
    \includegraphics[scale=1]{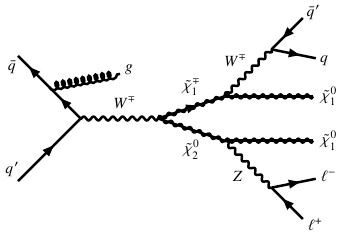}
    \caption{\label{fig:softLeptonGraph}Representative diagram for a $2\ell + E_{\text{T}}^{\text{miss}}$ signal from light electroweakino pair production. A $3\ell$ signal can arise from a leptonically decaying $W$ boson.}
\end{figure}

ATLAS-SUSY-2018-16 describes a search by the ATLAS Collaboration for electroweakino pair production in final states with missing transverse momentum and a low-momen\-tum opposite-sign same-flavor (OSSF) lepton pair \cite{ATLAS:2019lng}. This search is motivated by (and indeed targets) supersymmetric models in which the lightest neutralino $\tilde{\chi}^0_1$ is the LSP and is lighter than the next-lightest sparticles (\eg, electroweakinos $\{\tilde{\chi}^{\pm}_1, \tilde{\chi}^0_2\}$ or sleptons $\tilde{\ell}$) by only $\mathcal{O}(1\text{--}10)$~GeV. Such a spectrum allows a heavier state to produce a soft OSSF lepton pair through its decay to the LSP. A representative electroweakino signal process for this analysis is displayed in Figure~\ref{fig:softLeptonGraph}.

This analysis targets four simplified supersymmetric scenarios inspired by the MSSM. The \emph{higgsino} scenario features a triplet $\{\tilde{\chi}^0_1,\tilde{\chi}^{\pm}_1,\tilde{\chi}^0_2\}$ of light ``pure'' higgsinos with equal mass splittings $\Delta m(\tilde{\chi}^0_2,\tilde{\chi}^{\pm}_1)=  \Delta m(\tilde{\chi}^{\pm}_1,\tilde{\chi}^0_1)$, which resembles the MSSM electroweakino spectrum when $|\mu| \ll |M_1|,|M_2|$. The \emph{bino-wino} scenario, by contrast, is derived from the limit in which $|M_1| < |M_2| \ll |\mu|$ and features a mostly bino LSP lighter than a nearly degenerate pair of mostly wino $\{\tilde{\chi}^{\pm}_1,\tilde{\chi}^0_2\}$. The ATLAS search targets two bino-wino cases, motivated by the fact that the choice to make the neutralino mixing matrix real forces at least one mass eigenvalue to be negative: in the ``wino/bino(+)'' scenario, the two lightest neutralinos have masses of the same sign, while in the ``wino/bino($-$)'' scenario, the eigenvalues have opposite sign. We discuss the issue of signed electroweakino mass eigenvalues below, since it has physical implications and turns out to be important for the recasting of this analysis. Whereas in these first two scenarios, electroweakino production is assumed to be driven by $s$-channel $W/Z$ boson production, the third \emph{VBF} scenario targets $t$-channel electroweakino production through electroweakino exchanges between weak bosons. Finally, the \emph{slepton} scenario features the pair production of light sleptons, each decaying directly to the LSP and its lepton superpartner. 

\subsubsection{Selections}

\begin{table*}
\centering\renewcommand{\arraystretch}{1.2}\setlength{\tabcolsep}{8pt}
\begin{tabular}{l l l}
&\multicolumn{2}{c}{Preselection requirements}\\
\cline{2-3}\\[-1.0em]
Variable   		                     & $2\ell$  					     & $1\ell 1 T$\\
\midrule
Number of leptons (tracks)           & 2 leptons 								 & 1~lepton~and~$\geq1$~track\\
Lepton $p_{\text{T}}$~[GeV]      & $p_{\text{T}}(\ell_1)>5$                                   & $p_{\text{T}}(\ell)<10$				\\
$\Delta R_{\ell\ell}$             & $\Delta R_{ee}>0.30$, $\Delta R_{\mu\mu}>0.05$, $\Delta R_{e\mu}>0.2$\ \ \ \ \ \ \ \ \     & $0.05<\Delta R_{\ell,\mathrm{track}}<1.5$\\
Lepton (track) charge and flavor     & $e^{\pm}e^{\mp}$ or $\mu^{\pm}\mu^{\mp}$      & $e^{\pm}e^{\mp}$ or $\mu^{\pm}\mu^{\mp}$ \\
Lepton (track) invariant mass~[GeV]\ \ \ \ \ \ \ \ \ & $3<m_{ee}<60$, $1<m_{\mu\mu}<60$              		 & $0.5<m_{\ell,\mathrm{track}}<5$\\
$J/\psi$ invariant mass~[GeV]       & {\textsc{veto}} $3<m_{\ell\ell}<3.2$                             & {\textsc{veto}} $3<m_{\ell,\mathrm{track}}<3.2$\\
$m_{\tau\tau}$~[GeV]                    & $<0$ or $>160$                                & no requirement \\
$E_{\text{T}}^{\text{miss}}$~[GeV]                          & $>120$                                        & $>120$ \\
Number of jets                       & $\geq 1$                                      & $\geq 1$ \\
Number of $b$-tagged jets      	     & $=0$                                          & no requirement \\
Leading jet $p_{\text{T}}$~[GeV]               & $\geq 100$                                    & $\geq 100$\\
$\min(\Delta\phi(\mathrm{any\,jet},\tv{p}_{\text{T}}^{\text{miss}}))$             	     & $>0.4$                                        & $>0.4$ \\
$\Delta \phi(j_1,\tv{p}_{\text{T}}^{\text{miss}})$                          & $\geq 2.0$                                    & $\geq 2.0$ \\
\end{tabular}
\caption{Preselection requirements applied to all events entering into electroweakino search regions of ATLAS-SUSY-2018-16.}\label{table:signalreg:preselectioncuts}
\end{table*}

This analysis comprises several classes of signal region (SR), each targeting one or more of the BSM scenarios described above. The SRs are categorized as E (electroweakino), VBF, and S (slepton). While we have mentioned the latter two for completeness, only the electroweakino signal regions are recast; thus the rest of this discussion is limited to the signal regions labeled by the ATLAS Collaboration as SR--E (with various suffixes in the exclusive channels). The electroweakino search itself contains two channels: the $2\ell$ channel is characterized by a pair of OSSF signal leptons, whereas the $1\ell 1 T$ channel, which targets the smallest mass splittings, requires one lepton and at least one low-$p_{\text{T}}$ track that can be matched to a lepton of opposite sign.

The preselection requirements are listed in Table \ref{table:signalreg:preselectioncuts}. As we explore below, the excesses appear in the $2\ell$ channels. The leading lepton must have transverse momentum $p_{\text{T}} > 5$~GeV, and the lepton pair must satisfy various requirements on isolation and its invariant mass, depending on the OSSF pair flavor. Missing transverse energy (\MET) of at least 120~GeV must be recorded, and at least one jet with $p_{\text{T}} \geq 100$~GeV is required. There are finally isolation requirements on the missing transverse momentum and both the leading jet and any additional jets.

\begin{table*}
\centering\renewcommand{\arraystretch}{1.2}\setlength{\tabcolsep}{8pt}
\begin{tabular}{l l l l l }
&\multicolumn{4}{c}{Electroweakino SR Requirements}                                                                                 \\
\cline{2-5}\\[-1.0em]
Variable                           &SR--E--low           & SR--E--med            &SR--E--high                             &SR--E--$1\ell 1 T$ \\
\midrule
$E_{\text{T}}^{\text{miss}}$~[GeV]                        &$[120, 200]$         &$[120, 200]$           &$>200$                                  & $>200$                  \\
$E_{\text{T}}^{\text{miss}}/H_{\text{T}}^{\text{lep}}$                        &$<10$                &$>10$                  &--                                      & $>30$                \\
$\Delta \phi(\text{lep},\tv{p}_{\text{T}}^{\text{miss}})$	                       &--				   	 &--				     &--					                  & $<1.0$               \\
Lepton or track $p_{\text{T}}$~[GeV]\ \ \ \ \       & $p_{\text{T}}(\ell_2) > 5 + m_{\ell\ell}/4$\ \ \ \ \ &--                     &$p_{\text{T}}(\ell_2)>\min(10, 2 + m_{\ell\ell}/3)$          & $p_{\text{T}}^{\text{track}}<5$       \\
$M_{\text{T}}^{S}$~[GeV]                        &--                   &$<50$                  &--                                      &--                   \\
$m_{\text{T}}(\ell_1)$~[GeV]                     &$[10,60]$            &--                     &$<60$                                   &--                   \\
$R_{\text{ISR}}$                             &$[0.8,1.0]$          &--                     &$[\max(0.85,0.98-0.02\times m_{\ell\ell}),~1.0]$\ \ \ \ \ &--                   \\
\end{tabular}
\caption{Selection requirements for the four electroweakino signal regions in ATLAS-SUSY-2018-16.}
\label{table:signalreg:ewkinocuts}
\end{table*}

\begin{table*}\renewcommand{\arraystretch}{1.1}\setlength{\tabcolsep}{5pt}
\begin{tabular}{c l rrrrrrrr }
\ \ \ \ \ \ \ \ \ \ \ \ \ \ \ \ \ & \ \ \ \ \ \ \ \ \ \ \ \ \ \ \ \ \ \ \ \ \ \ \ \ \ \ \ \ \ \ \ & \multicolumn{8}{c}{SR $m_{\ell\ell}$ bin [GeV]}\\
\cline{3-10}\\[-1.0em]
& &{[1,2]} & {[2,3]} &{[3.2,5]} & {[5,10]} & {[10,20]} & {[20,30]} & {[30,40]} & {[40,60]} \\
\midrule
\multirow{2}{*}[-0.5ex]{\makecell{SR--E\\ high $ee$}} &Observed  &&& $1$ & {\color{blue}\textbf{16}} & $13$ & {\color{red}8} & {\color{red}8} & $18$ \\
\cmidrule{2-10}
\cmidrule{2-10}
&Fitted SM events &&& $0.7 \pm 0.4$ & $10.3 \pm 2.5$ & $12.1 \pm 2.2$ & $10.1 \pm 1.7$ & $10.4 \pm 1.7$ & $19.3 \pm 2.5$ \\
\midrule
\multirow{2}{*}[-0.5ex]{\makecell{SR--E\\ high $\mu\mu$}}
&Observed & {\color{blue}5} & {\color{blue}5} & {\color{red}\emph{0}} & $9$ & {\color{blue}23} & {\color{red}\emph{3}} & {\color{red}\emph{5}} & $20$ \\
\cmidrule{2-10}
\cmidrule{2-10}
&Fitted SM events & $3.4 \pm 1.2$ & $3.5 \pm 1.3$ & $3.9 \pm 1.3$ & $11.0 \pm 2.0$ & $17.8 \pm 2.7$ & $8.3 \pm 1.4$ & $10.1 \pm 1.5$ & $19.6 \pm 2.3$ \\
\midrule
\multirow{2}{*}[-0.5ex]{\makecell{SR--E\\ med $ee$}}
&Observed  &&& $0$ & $4$ & {\color{blue}11} & {\color{blue}\textbf{4}} \\
\cmidrule{2-10}
\cmidrule{2-10}
&Fitted SM events &&& $0.11 \pm 0.08$ & $5.1 \pm 1.6$ & $7.3 \pm 1.9$ & $2.2 \pm 0.9$ \\
\midrule
\multirow{2}{*}[-0.5ex]{\makecell{SR--E\\ med $\mu\mu$}}
&Observed & $16$ & $8$ & $6$ & {\color{blue}41} & $59$ & $21$ \\
\cmidrule{2-10}
\cmidrule{2-10}
&Fitted SM events & $14.6 \pm 2.9$ & $6.9 \pm 2.1$ & $6.2 \pm 1.9$ & $34 \pm 4$ & $52 \pm 6$ & $18.5 \pm 3.2$ \\
\midrule
\multirow{2}{*}[-0.5ex]{\makecell{SR--E\\ low $ee$}}
&Observed &&& {\color{blue}7} & {\color{blue}11} & $16$ & $16$ & {\color{red}10} & {\color{red}\emph{9}} \\
\cmidrule{2-10}
\cmidrule{2-10}
&Fitted SM events &&& $5.3 \pm 1.5$ & $8.6 \pm 1.8$ & $16.7 \pm 2.5$ & $15.5 \pm 2.6$ & $12.9 \pm 2.1$ & $18.8 \pm 2.2$ \\
\midrule
\multirow{2}{*}[-0.5ex]{\makecell{SR--E\\ low $\mu\mu$}}
&Observed & {\color{red}\emph{9}} & $7$ & $7$ & $12$ & $17$ & $18$ & $16$ & {\color{blue}\textbf{44}} \\
\cmidrule{2-10}
\cmidrule{2-10}
&Fitted SM events & $15.4 \pm 2.4$ & $8.0 \pm 1.7$ & $6.5 \pm 1.6$ & $11.3 \pm 1.9$ & $15.6 \pm 2.3$ & $16.7 \pm 2.3$ & $15.3 \pm 2.0$ & $35.9 \pm 3.3$ \\
\end{tabular}
\caption{Observed event yields and fit results using a control region (CR) + SR background-only fit for the exclusive electroweakino $2\ell$ signal regions. Information on CRs, backgrounds, and uncertainties is available in the ATLAS analysis. Bins with upward fluctuations of greater than $1\sigma$ over the SM expectation are in {\color{blue}blue}, and those with excesses of $2\sigma$ or more are additionally in \textbf{bold}; bins with correspondingly large under-fluctuations are in {\color{red}red} and \emph{italics}.}
\label{tab:yields:higgsino:all}
\end{table*}

\begin{table*}
\centering\renewcommand{\arraystretch}{1.1}\setlength{\tabcolsep}{8pt}
\begin{tabular}{l rrrrrr }
\ \ \ \ \ \ \ \ \ \ \ \ \ \ \ \ \ \ \ \ \ \ \ \ \ \ \ \ \ \ \ & \multicolumn{6}{c}{SR $m_{\ell,\mathrm{track}}$ bin [GeV]}\\
\cline{2-7}\\[-1.0em]
 & {[0.5,1.0]} & {[1.0,1.5]} & {[1.5,2.0]} & {[2.0,3.0]} & {[3.2,4.0]} & {[4.0,5.0]} \\
\toprule
Observed & $0$ & $8$ & $8$ & $24$ & $24$ & $16$ \\
\midrule
Fitted SM events & $0.5 \pm 0.5$ & $6.0 \pm 1.9$ & $7.6 \pm 2.1$ & $20.7 \pm 3.4$ & $24 \pm 4$ & $18.1 \pm 3.1$ \\
\end{tabular}
\caption{Observed event yields and fit results using a CR + SR background-only fit for the exclusive electroweakino $1\ell 1 T$ regions. Information on CRs, backgrounds, and uncertainties is available in the ATLAS analysis. Only the $m_{\ell,\mathrm{track}} \in [1.0,1.5]~$GeV bin exhibits (barely) a $1\sigma$ upward fluctuation.}
\label{tab:1l1Tyields}
\end{table*}

\begin{table}
\centering
\renewcommand{\arraystretch}{1.1}\setlength{\tabcolsep}{6pt}
\begin{tabular}{l l r@{~$\pm$~\hspace{0ex}}l l}
Inclusive SR-E\ \ \ \ \ & $N_{\mathrm{obs}}$ & \multicolumn{2}{c}{$N_{\mathrm{exp}}$} & $p(s=0)$ \\
\midrule
$m_{\ell\ell}<1$~GeV    &   $0$   &   $1.0$ &$1.0 $\ \ \ \ \  & $0.50$   \\%
$m_{\ell\ell}<2$~GeV    &   $46$   &   $44$ &$6.8 $     & $0.38$   \\%
$m_{\ell\ell}<3$~GeV    &   $90$   &   $77$ &$12 $      & $0.18$   \\%
$m_{\ell\ell}<5$~GeV    &   $151$   &   $138$ &$18 $     & $0.24$   \\%
$m_{\ell\ell}<10$~GeV    &   $244$\ \ \ \ \   &   $200$ &$19 $     & {\color{blue}$\boldsymbol{0.034}$}   \\%
$m_{\ell\ell}<20$~GeV    &   $383$   &   $301$ &$23 $      & {\color{blue}$\boldsymbol{0.0034}$}   \\%
$m_{\ell\ell}<30$~GeV    &   $453$   &   $366$ &$27 $      & {\color{blue}$\boldsymbol{0.0065}$}   \\%
$m_{\ell\ell}<40$~GeV    &   $492$   &   $420$ &$30 $   & {\color{blue}$\boldsymbol{0.027}$}   \\%
$m_{\ell\ell}<60$~GeV    &   $583$   &   $520$ &$35 $  & $0.063$   \\%
\end{tabular}
\caption{Observed and expected event yields for the inclusive electroweakino regions, defined as the union of the individual electroweakino SRs (both $2\ell$ and $1\ell 1 T$ channels) for specified upper bounds on $m_{\ell\ell}$. Information on background predictions and uncertainties is available in the ATLAS analysis. A discovery $p$-value ($p(s = 0)$) is computed for each inclusive bin. The four inclusive bins with a $p$-value smaller than 0.05 are in {\color{blue}blue} and \textbf{bold}.}
\label{tab:Upperlimit_SRSF_allmet}
\end{table}
\renewcommand{\arraystretch}{1.0}

The selection requirements for the electroweakino signal regions are listed in Table \ref{table:signalreg:ewkinocuts}. As noted above, the electroweakino analysis includes $2\ell$ and $1\ell 1 T$ channels. The $2\ell$ channel moreover contains exclusive signal regions distinguished by requirements on $E_{\text{T}}^{\text{miss}}$ and the ratio $E_{\text{T}}^{\text{miss}}/H_{\text{T}}^{\text{lep}}$ of missing transverse energy to the scalar sum of lepton $p_{\text{T}}$. (The exclusive analysis also reports yields separately for different OSSF lepton pair flavors.) In addition to the $E_{\text{T}}^{\text{miss}}$ requirements and additional lepton $p_{\text{T}}$ and isolation cuts, the electroweakino selection imposes requirements on two variables intended to select events with high $E_{\text{T}}^{\text{miss}}$ resulting from electroweakino recoils off of initial state radiation (ISR). These variables are defined in terms of hemispheres constructed using the recursive jigsaw reconstruction (RJR) technique \cite{Jackson:2017gcy} orthogonal to the \emph{thrust axis} along which electroweakinos recoil against ISR. They are $M_{\text{T}}^{S}$, the transverse mass of the objects in the \emph{supersymmetric-particles hemisphere} $S$; and $R_{\text{ISR}}$, the ratio of $E_{\text{T}}^{\text{miss}}$ to the transverse momentum of the objects in the \emph{ISR hemisphere}. A requirement on $M_{\text{T}}^S$ is imposed on SR--E--med, while $R_{\text{ISR}}$ cuts are placed on SR--E--low and SR--E--high. 

In the exclusive analysis, once the cuts in Table \ref{table:signalreg:ewkinocuts} are imposed, the dilepton invariant mass ($m_{\ell\ell}$ in the $2\ell$ channel and $m_{\ell,\text{track}}$ in the $1\ell 1 T$ channel) is binned between 1~GeV and up to 60~GeV for the $2\ell$ channel and between 0.5~GeV and 5.0~GeV for the $1\ell 1 T$ channel, which again targets BSM spectra with quite strong compression. An inclusive electroweakino analysis is also performed by merging all $2\ell$ and $1\ell 1 T$ bins below an array of $m_{\ell\ell}$ from 1~GeV to 60~GeV. A ``model-independent'' discovery $p$-value is then computed for each bin in the inclusive analysis, and a model-dependent exclusion fit is performed for the exclusive bins.

\subsubsection{\texorpdfstring{$m_{\ell\ell}$}{m\_ell\_ell} distributions: Anatomy of the excess}

The expected and observed yields in the exclusive $m_{\ell\ell}$ and $m_{\ell,\text{track}}$ bins are displayed in Table \ref{tab:yields:higgsino:all} for the $2\ell$ channels and in Table \ref{tab:1l1Tyields} for the $1\ell 1 T$ channel. The binning of the signal regions according to $m_{\ell\ell}$ allows for better discrimination between signal and background, and this is therefore the strategy also employed by the equivalent CMS analysis \cite{CMS:2021edw}. In principle it also affords a discrimination between different supersymmetric scenarios, because the distribution of the decays is different between the higgsino and wino/bino(+) cases. The excess is noticeable in several $2\ell$ bins, mostly with moderate $m_{\ell\ell}$ between 5~GeV and 20~GeV, and the accumulation of excess events in these exclusive bins is reflected in the significant discovery $p$-values computed by the ATLAS Collaboration for four inclusive signal regions with $m_{\ell \ell} < 40$~GeV, as shown in Table \ref{tab:Upperlimit_SRSF_allmet}. However, it is notable that there are significant \emph{under}fluctuations in several exclusive bins: those of $[3.2,5], [20,30], [30,40]$~GeV for the high-\MET $\mu\mu$ region, the $[40,60]$~GeV bin for the low-\MET $ee$ region, and the $[1,2]$~GeV bin for the low-\MET $\mu\mu$ region. These are potentially problematic for any model attempting to explain the excess: because the underfluctuations are so severe, predicting even a few events in these bins would rule out the model. This means that models that predict large numbers of events with $m_{\ell\ell} > 20$ GeV or $m_{\ell\ell} < 5$ GeV are likely excluded, and (as we shall see) means that it is possible to obtain exclusion limits \emph{stronger} than the expected ones for such cases. Hence models that best fit the excess should have an $m_{\ell\ell}$ distribution peaked in the $[5,10]$ GeV range, and should produce both electron and muon pairs, though seemingly not with identical kinematics. 

\begin{figure*}
    \centering
    \includegraphics[scale=0.72]{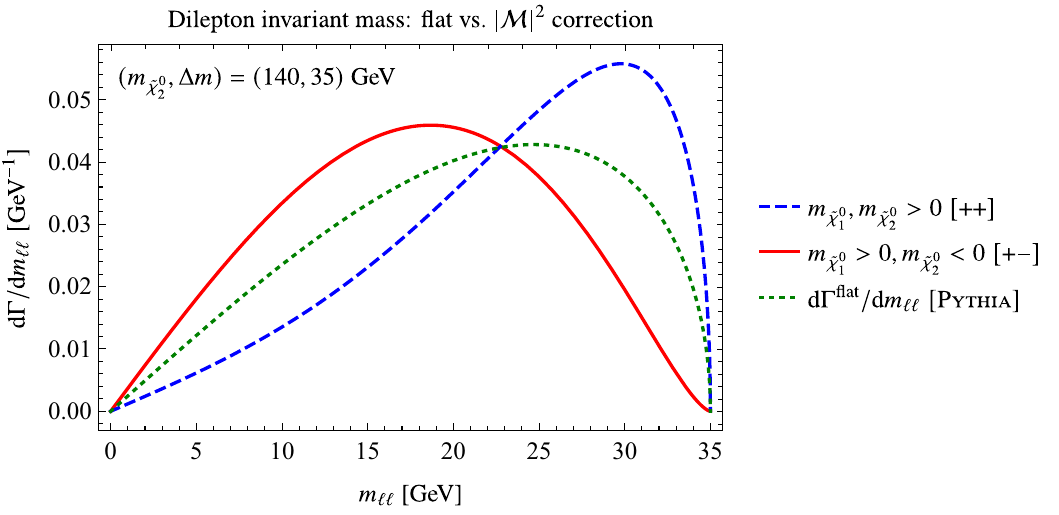}
    \caption{Dilepton invariant mass distributions for the three cases: both neutralino eigenvalues positive (blue, dashed); opposite-sign eigenvalues (red); flat phase space distribution in which the associated squared matrix element $|\mathcal{M}|^2$ is set to 1 (green, dotted).}\label{fig:mllcomparison}
\end{figure*}

In the ATLAS (and equivalent CMS) publications, the simplified scenarios considered both produce dilepton events emerging from the production of heavy neutralinos that decay to the LSP via an off-shell $Z$ boson, $\tilde{\chi}_2^0 \rightarrow Z^* +\tilde{\chi}_1^0.$ The $m_{\ell \ell}$ distributions of the dileptons produced in this manner are thus invariant under boosts and depend only on the couplings of the neutralinos and the mass differences. The full differential decay rate, ignoring the masses of the leptons, is given by~\cite{Montesano:2005pga,DeSanctis:2006tbe}
\begin{multline}\label{eq:dGdmll}
\frac{\d\Gamma}{\d m_{\ell\ell}} = C m_{\ell\ell}\, \frac{\{m_{\ell\ell}^4 - m_{\ell\ell}^2 [(\Delta m)^2 + M^2] + (M \Delta m )^2\}^{1/2}}{(m_{\ell\ell}^2 - M_Z^2)^2}\\ \times \{ - 2 m_{\ell\ell}^4 + m_{\ell\ell}^2 [2M^2 - (\Delta m)^2] + (M \Delta m)^2\},
\end{multline}
where $\Delta m = m_{\tilde{\chi}_2^0} - m_{\tilde{\chi}_1^0}$ and $M = m_{\tilde{\chi}_2^0} + m_{\tilde{\chi}_1^0}$ in the basis where the mixing matrices are real: these are \emph{signed eigenvalues}. $C$ represents a normalization constant. Note that there is a sign error in \cite{Montesano:2005pga,DeSanctis:2006tbe} that we correct in \eqref{eq:dGdmll}.\footnote{In Equation (4) of \cite{DeSanctis:2006tbe}, this comes from (4.21) in \cite{Montesano:2005pga}, but the previous formula (4.18) is correct.} Hence if they are of opposite sign, with $m_{\tilde{\chi}_2^0}$ negative as in the ATLAS higgsino and wino/bino($-$) scenarios, then $\Delta m$ and $M$ swap roles; the difference stems from the $2M^2 - (\Delta m)^2$ term.

The expression \eqref{eq:dGdmll} vanishes at zero and the minimum of $m= |\Delta m|, |M|$. By contrast, in the case of a flat phase space (as used in typical {\tt SLHA} decay processes in \pythia, where the contribution from the squared matrix element $|\mathcal{M}|^2$ is unknown and therefore set to unity) the differential decay rate reduces to
\begin{multline}
\frac{\d\Gamma^{\rm flat}}{\d m_{\ell\ell}} = C^{\rm flat}\, m_{\ell\ell}\\ \times \{m_{\ell\ell}^4 - m_{\ell\ell}^2 [(\Delta m)^2 + M^2] + (M \Delta m)^2\}^{1/2}.
\end{multline}
We plot these three cases in Figure \ref{fig:mllcomparison} in a scenario where the NLSP-LSP splitting is $\Delta m = 35$~GeV. While it is clear that a flat distribution does not discriminate between the two signed-eigenvalue scenarios, it turns out that this is not a mere curiosity and the accurate $m_{\ell\ell}$ distributions are vitally important to obtaining a successful reproduction of the experimental results, and therefore also to further interpretations. 

In the simplified higgsino scenario (where the eigenvalues have opposite signs), we see that the distribution is flatter and peaked near $\Delta m/2$; whereas in the wino/bino(+) scenario (where they have the same sign) the peak is much closer to the upper endpoint. Hence if we seek to explain an excess by producing events in the range $m_{\ell \ell} \in [5,20]$ GeV we expect the best fit for the excess for the higgsino case to be around $\Delta m \in [10,40] $ GeV, and for the wino/bino(+) case to be roughly in the region $\Delta m \in [6, 24]$ GeV. This explains the features in the exclusion plots; therefore, in other models we can seek to explain the excess (or lack thereof) by plotting their $m_{\ell \ell}$ distributions, as we do through the rest of this work.

\subsubsection{Recast and validation}

This analysis presents several challenges for both implementation and validation. In the following we describe how this has been achieved in \hackanalysis v2, which is now publicly available. This was possible because the ATLAS Collaboration provided substantial supplementary material: a pseudocode  for {\sc SimpleAnalysis}~\cite{ATLAS:2022yru}; a \pyhf~\cite{Heinrich:2021gyp} model file for the background model; a set of \emph{signal} patches for the \pyhf model file corresponding to a wide variety of higgsino and wino/bino signal points; and cutflows for one parameter point (the higgsino case with $m_{\tilde{\chi}_2^0} = 155$~GeV, $\Delta m = 5$~GeV).

Of the major recasting frameworks, only {\sc CheckMATE} \cite{Drees:2013wra,Dercks:2016npn} has an extant version of this analysis, although seemingly the only validation that was performed was a comparison with the cutflows provided by ATLAS. The main difficulty with recasting this analysis is that it relies on RJR variables (\emph{viz}. Section~\ref{s2.1}) constructed using the {\sc RestFrames} package \cite{restframes}; this is a large package that relies on {\tt root}~\cite{Brun:1997pa} and many of its functionalities (especially rotations of vectors and minimization methods). Since {\sc GAMBIT}~\cite{GAMBIT:2017yxo} and \madanalysis~\cite{Conte:2012fm,Conte:2014zja,Conte:2018vmg} use their own treatments of Lorentz vectors and other features of {\tt root}, it is not simple to link {\sc Restframes} within these platforms. On the other hand, {\sc CheckMATE} does use {\tt root} objects, but in their implementation of this analysis only pseudo-jigsaw variables are used as a proxy to the full use of {\sc RestFrames} and the package is not linked. 

However, \hackanalysis was designed to be easily hackable, and so it was possible to create a {\tt root}-free implementation of {\sc RestFrames} within \hackanalysis. This required the addition of new features in \hackanalysis to replace those required from {\tt root}: in particular, minimization routines based on the Nelder-Mead algorithm~\cite{Nelder:1965zz}; and matrix manipulations now provided by the {\tt Eigen} package.\footnote{See the webpage \url{https://gitlab.com/libeigen/eigen}.} It was possible, however, to adapt the {\sc SimpleAnalysis} helper functions that abstract the interface to {\sc RestFrames} such that they could be called like in the pseudocode, and this enables straightforward implementation of any other analyses requiring {\sc RestFrames}. The new functions available in \hackanalysis v2 are detailed in the documentation with the code, and the new features are described in a separate manual \cite{Goodsell:2024aig}.

Efficiencies for the reconstruction of the signal leptons and tracks were provided in the form of plots, from which it was necessary to scrape the data; these were implemented in the form of functions fit to the scraped data. However, as shown in Table \ref{table:signalreg:ewkinocuts}, the lowest cut on missing energy in the analysis is $E_{\text{T}}^{\text{miss}} \geq 120$ GeV, where it is known that the trigger efficiency is not high; this requires a reweighting of the events based on scraped data on the ATLAS \MET trigger efficiency found in dedicated ATLAS searches. With the efficiencies in hand, it becomes possible to validate the analysis in several ways. We performed a comparison of the cutflows, a comparison of the final yields with those in the signal patchset, and a reproduction of the exclusion plots. We note that the pseudocode provided contains only signal regions (not control regions); using the patchsets, we were able to confirm that the signal contamination in control regions is small (especially compared to the backgrounds) and therefore we only implemented the signal regions in the recast. 

One of the striking features of this analysis is its diminutive selection efficiencies: typical cross sections for electroweakino production in the mass range to which this analysis is sensitive are of $\mathcal{O}(1)$~pb, yet producing even three events in the wrong $m_{\ell\ell}$ bin would exclude a model. We thus find typical efficiencies of $\mathcal{O}(10^{-5})$ and for many cases (especially at lower masses and lower $\Delta m$) even smaller. Such low efficiencies demand quite large samples in order to achieve acceptable numerical uncertainty and therefore make the simulation of the signals technically challenging.

\begin{figure*}
    \centering
    \includegraphics[scale=0.7]{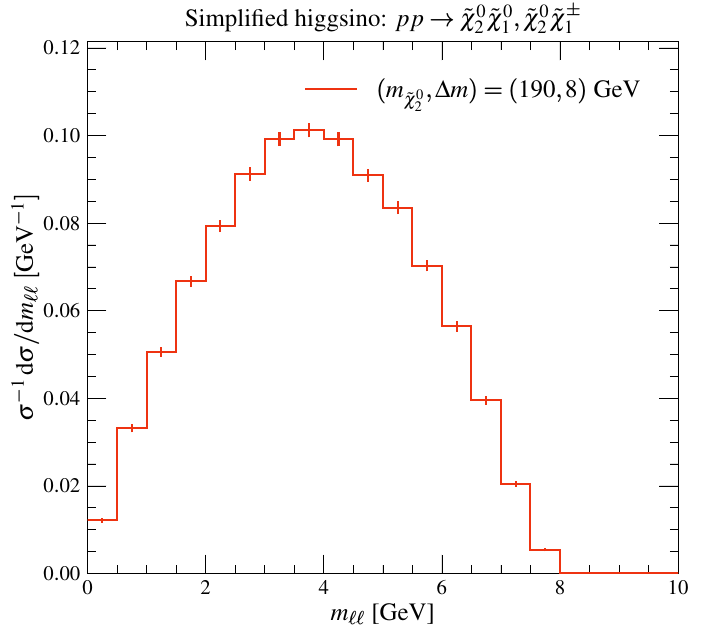}\hspace{1cm}\includegraphics[scale=0.7]{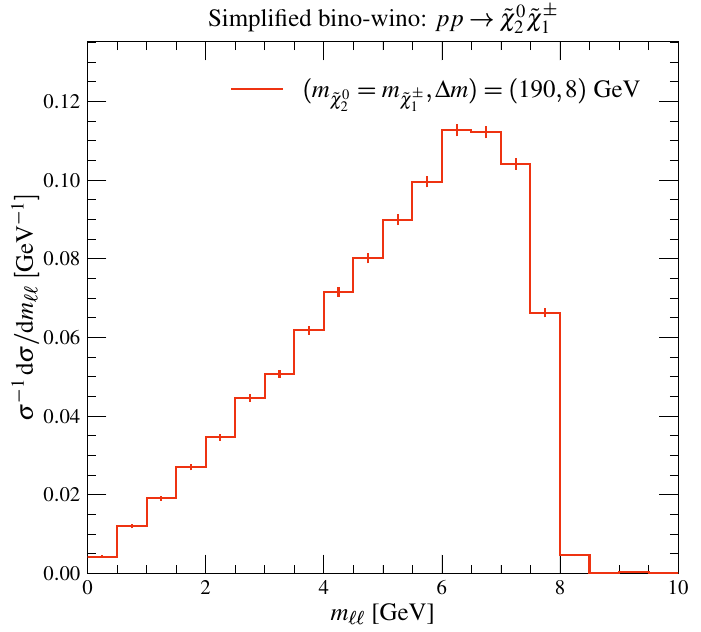}
    \caption{$m_{\ell\ell}$ distributions for the ATLAS simplified higgsino (left) and bino-wino [wino/bino(+)] (right) cases as found from our simulations for the processes $p p \rightarrow \tilde{\chi}_2^0\tilde{\chi}^{\pm}_1$ (and $pp \to \tilde{\chi}^0_2\tilde{\chi}^0_1$ for higgsinos) with matrix elements including up to two additional partons.}\label{fig:comparemll}
\end{figure*}

\begin{figure}
    \centering
    \includegraphics[scale=0.7]{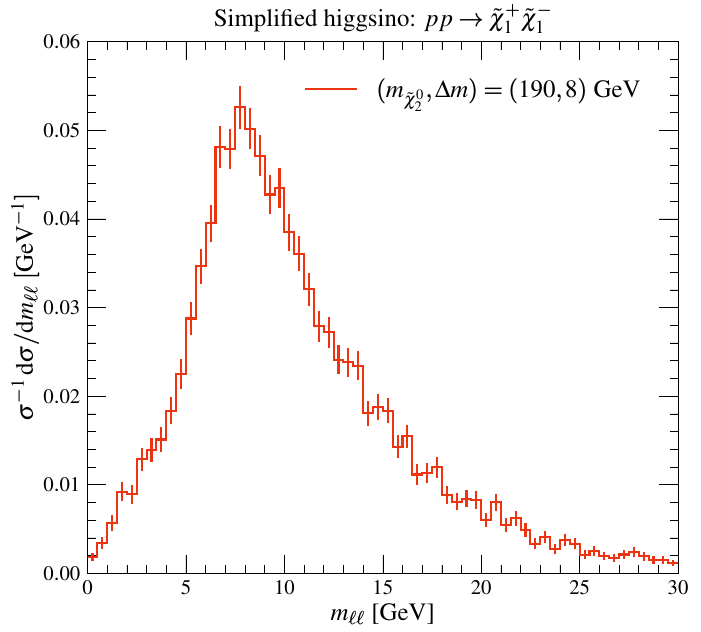}
    \caption{$m_{\ell\ell}$ distribution for the ATLAS simplified higgsino case as found from our simulations for the chargino pair-production process $p p \rightarrow \tilde{\chi}_1^+ \tilde{\chi}_1^-$, with matrix elements including up to two additional partons.}\label{fig:higgsinoc1c1mll}
\end{figure}

In the discussion above, we emphasized the importance of the $m_{\ell\ell}$ distributions due to the use of this variable to create signal bins, and because there are severe underfluctuations in certain bins that lead to strong limits upon any model that populates those bins. In the validation we found this to be particularly striking: it is impossible to validate this analysis using a flat phase-space distribution of the three-body decays of neutralinos. To obtain the correct distributions we could, in principle, use {\sc MadSpin}~\cite{Artoisenet:2012st} and {\sc MadWidth}~\cite{Alwall:2014bza} if we prepared different event samples for each final state. Instead, we use the decay chain functionality of \textsc{MadGraph5\texttt{\textunderscore}aMC@NLO} (\MadGraph) \cite{Alwall:2014hca} to compute the decays including full spin correlations via commands such as
\begin{align*}
    \texttt{MG5\textunderscore aMC> generate p p > n1 n2, n2 > n1 l+ l-}.
\end{align*}
In this way we force the heavier neutralino to decay to leptons and gain an order of magnitude of efficiency at the detector level, because this branching ratio is less than $\mathcal{O}(10)\%.$ On the other hand, for processes involving charginos, since only one lepton can come from their decays, the exact distribution is less important: the reconstructed $m_{\ell \ell}$ will depend more on the mother particle momenta. Hence we model the decays of charginos with a flat phase space in \pythia~\cite{Bierlich:2022pfr}.

Even with such improvement, to obtain uncertainties of $\mathcal{O}(10)\%$ on efficiencies of $\mathcal{O}(10^{-5})$ requires $\mathcal{O}(10^6)$ events. Moreover, for ATLAS' higgsino model we need to separately simulate both $pp \to \tilde{\chi}_2^0 \tilde{\chi}_1^\pm,\tilde{\chi}_2^0\tilde{\chi}_1^0$ \emph{and} $pp \to \tilde{\chi}^+_1 \tilde{\chi}^-_1$. Even further, in order to obtain the correct distributions for the missing energy, which depends on the recoil of the electroweakinos from initial-state radiation, we must combine matrix elements including up to two additional partons. After matching the matrix elements in the MLM scheme \cite{Mangano:2006rw, Alwall:2008qv} with a matching scale $Q_{\mathrm{cut}}$ fixed\footnote{A typical value for the matching scale is $Q_{\mathrm{cut}} = m_{\tilde{\chi}}/4$ or the average mass of the produced electroweakinos.} at $40$~GeV, we are left with about half of the simulated hard-scattering events. In order to meet these challenges, we leverage the ability to run \hackanalysis on multiple cores: using a new gridpack mode within \BSMArt \cite{Goodsell:2023iac}, we generate for each point in the relevant parameter space a gridpack from \MadGraph, involving the built-in UFO~\cite{Degrande:2011ua, Darme:2023jdn} MSSM implementation~\cite{Duhr:2011se} and the leading-order set of NNPDF2.3 parton densities~\cite{Ball:2012cx, Buckley:2014ana}, and use it to generate one Les Houches Event file~\cite{Alwall:2006yp} for each core, running in parallel. This eliminates one of the major bottlenecks in \MadGraph (at least for versions prior to 3.5.4) where the simulated events generated for each subprocess are only recombined and reweighted using a single core. Then parton showering is performed within \hackanalysis using \pythia~\cite{Bierlich:2022pfr} so that no showered events need be written to disk. In this way, using eight cores, we can analyze $3.2 \times 10^6$ events in around four hours per parameter point. We simulate $\mathcal{O}(10^2)$ parameter points for each model and each analysis in this work; the number of points is specified in the discussion of each model.

We book histograms of the $m_{\ell\ell}$ distributions in \hackanalysis using \YODA \cite{Buckley:2023xqh}; in Figure~\ref{fig:comparemll} we plot the distributions (using the convenient plotting facility of \YODA) obtained for the process $p p \rightarrow \tilde{\chi}_2^0 +X$, for the higgsino and wino/bino(+) scenarios, for the parameter point $m_{\tilde{\chi}_2^0} = 190$ GeV, $\Delta m = 8$~GeV. We find that the generated distributions match the predictions in Figure~\ref{fig:mllcomparison}. On the other hand, a stark difference is visible in Figure~\ref{fig:higgsinoc1c1mll} for the case of (simplified higgsino) chargino pair production, in which each lepton in the OSSF pair comes from the chargino of matching charge. This distribution is sharply peaked at $\Delta m$ but has a comparatively long tail for larger $m_{\ell\ell}$. This result suggests that models with decays purely via off-shell $W$ boson(s) will be harder, though not necessarily impossible, to prevent from generating too many events in the larger $m_{\ell \ell}$ bins.

\begin{figure*}
    \centering
    \includegraphics[width=0.47\textwidth]{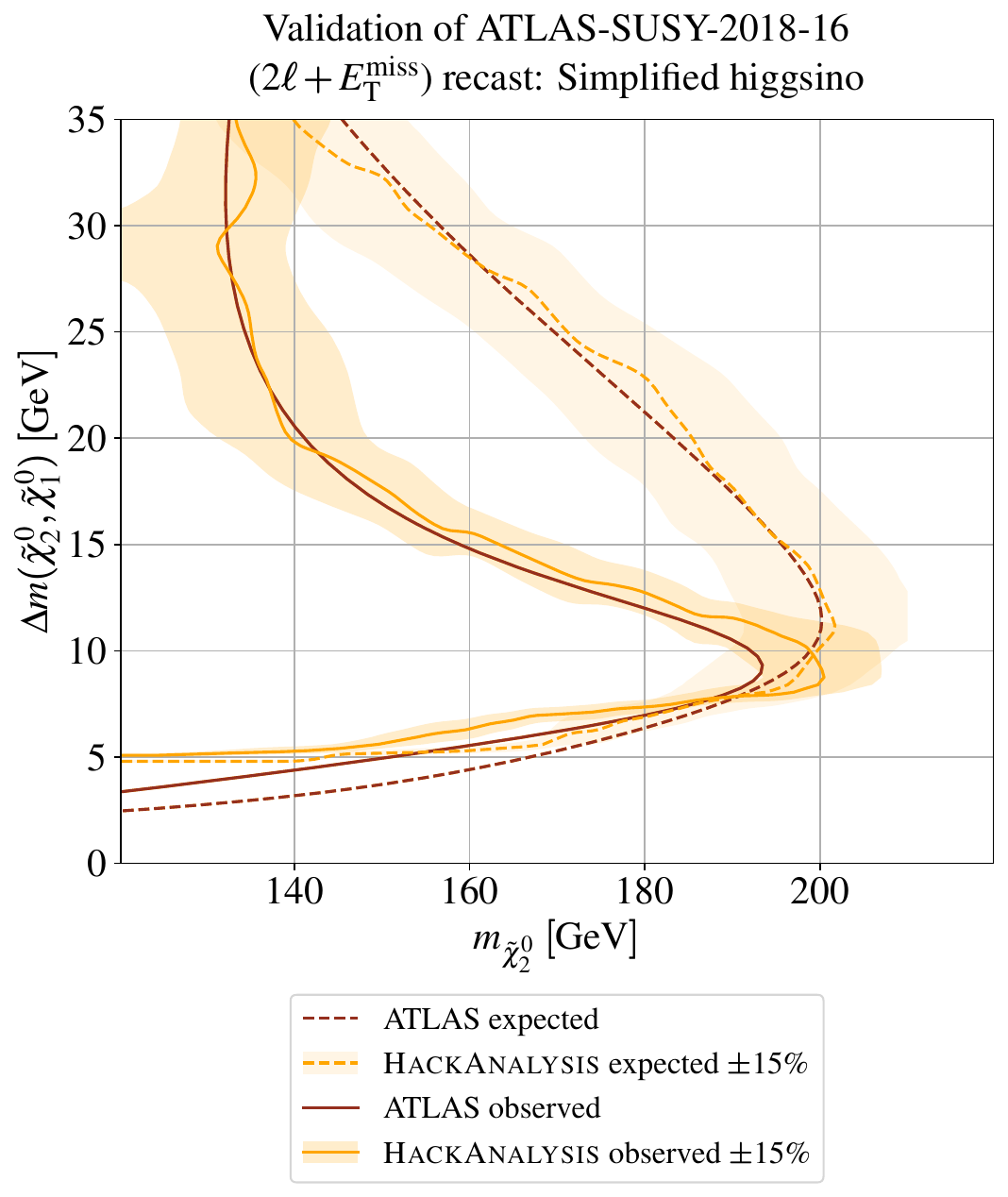}\hfill
    \includegraphics[width=0.47\textwidth]{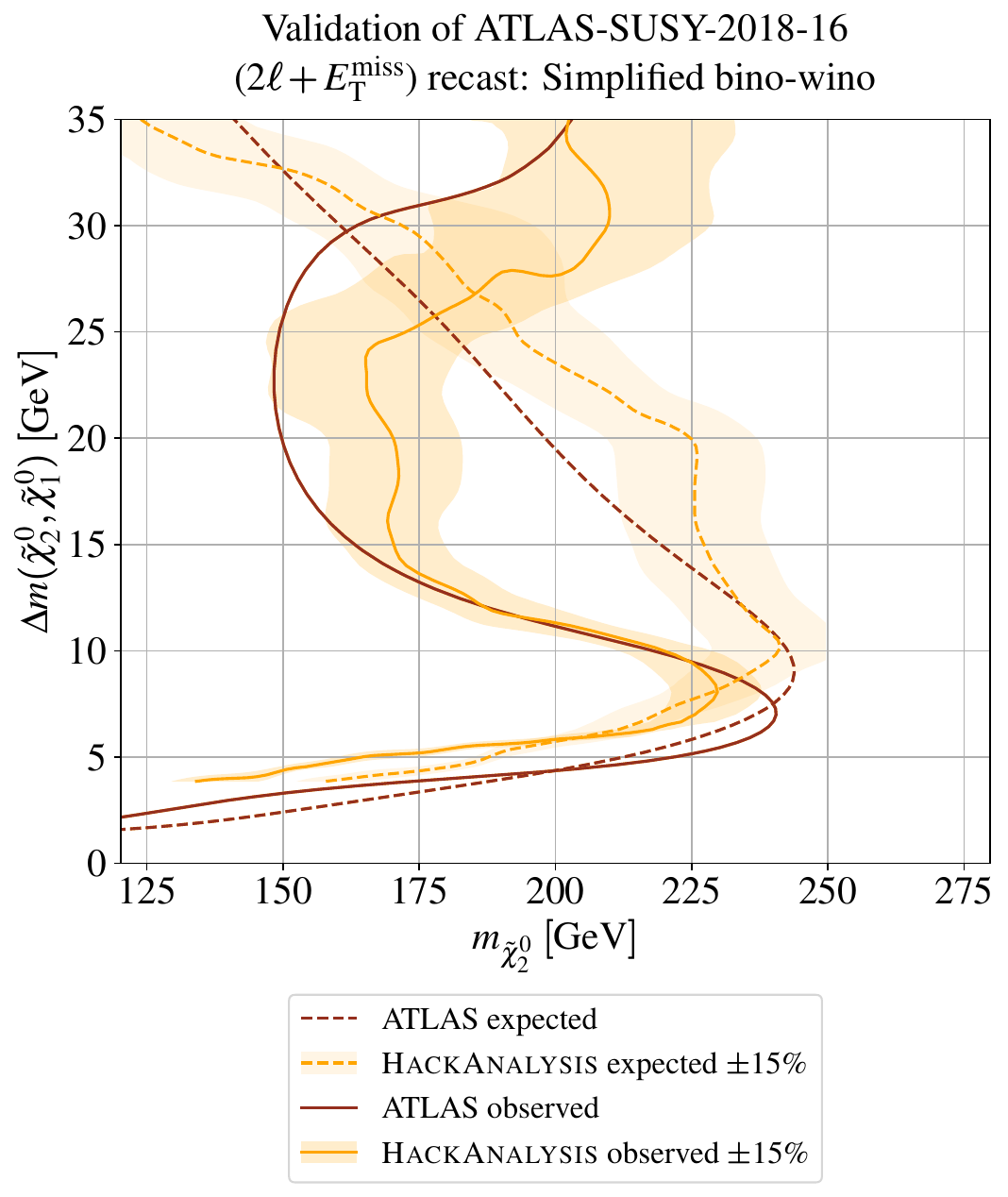}
    \caption{Comparison of \textsc{HackAnalysis} exclusions to official ATLAS results for ATLAS-SUSY-2018-16. Left panel shows exclusions in the simplified higgsino model while right panel refers to the simplified bino-wino scenario. \textsc{HackAnalysis} uncertainty bands correspond to 15\% variation of the signal strength.}\label{fig:2lValid}
\end{figure*}

To validate our recast, after comparing the cutflows for the one higgsino point given by ATLAS, we compare signal yields with selected points from the patchset and find excellent agreement. A reliable comparison requires signal cross sections at next-to-leading order in the strong coupling, including next-to-leading logarithmic corrections (NLO + NLL), which we compute for each of our parameter points using {\sc Resummino}~\cite{Fuks:2013vua,Fiaschi:2023tkq}. However, in the patchset it is also evident that the provided signals contain a \emph{statistical uncertainty}: the numbers simulated by ATLAS include uncertainties of about 5\% on the most sensitive signal regions and larger uncertainties on those with smaller number of events. Accounting for this uncertainty makes it possible to validate our recast with exclusion plots for the simplified higgsino and bino-wino cases (specifically wino/bino(+) in the latter case). We simulate 227 simplified higgsino and 368 simplified bino-wino points for this validation. In reproducing ATLAS' exclusion plots, we find that the statistical uncertainty weakens the observed limits by around $20$~GeV; it is striking that ATLAS could presumably have improved the reach of their search by simply simulating more signal events. We therefore show a comparison of the expected and observed limits for the simplified higgsino and bino-wino cases in Figure~\ref{fig:2lValid} using a flat uncertainty in the \pyhf patch of $5\%$. To be conservative, we further include bands corresponding to downward and upward 15\% variations of the signal cross sections. Altogether, we find excellent agreement for both models.
 
\subsection{ATLAS-SUSY-2019-09: Three leptons + missing energy}

ATLAS-SUSY-2019-09 is described by the ATLAS Collaboration as a search for electroweakino pair production in final states with two or three leptons and missing transverse momentum \cite{ATLAS:2021moa}. More precisely, it consists of a search in the $3\ell + E_{\text{T}}^{\text{miss}}$ channel, which is statistically combined with the results of ATLAS-SUSY-2018-16 in the $2\ell + E_{\text{T}}^{\text{miss}}$ channel, discussed in the previous subsection. Taken together, these searches target pair production of light MSSM electroweakinos with mass splittings ranging from quite small values ($\mathcal{O}(1)~\text{GeV}$) to over 100~GeV, depending on the model. To avoid confusion, since we do not perform any statistical combination with our recasts, we clarify here that we use the analysis identifier ``ATLAS-SUSY-2019-09'' to refer specifically to the $3\ell$ channel of the combined analysis. The most important process in the $3\ell$ channel is production of the lightest chargino and the second-lightest neutralino, $pp \to \tilde{\chi}^{\pm}_1\tilde{\chi}^0_2$. The chargino is again assumed to decay to the LSP and a $W$ boson in all cases, though as before the $W$ may be off shell. Final states with three leptons arise from leptonic decays of the $W$, so that Figure~\ref{fig:softLeptonGraph} is easily altered to show a representative $3\ell$ signal process. Meanwhile, the assumed decay of $\tilde{\chi}^0_2$ is always to the LSP and an opposite-sign lepton ($\ell \in \{e,\mu\}$) pair, but the SM boson assumed to mediate this decay (and whether it is on shell) determines the selection criteria and the interpretation of the results. Namely, ATLAS defines the following selections:
\begin{itemize}
    \item \textbf{On-shell $\boldsymbol{WZ}$} selection: $\tilde{\chi}^0_2 \to Z\tilde{\chi}^0_1$,
    \item \textbf{Off-shell $\boldsymbol{WZ}$} selection: $\tilde{\chi}^0_2 \to Z^* \tilde{\chi}^0_1$,
    \item \textbf{$\boldsymbol{Wh}$} selection: $\tilde{\chi}^0_2 \to h \tilde{\chi}^0_1$.
\end{itemize}
In this work we only make use of the off-shell $WZ$ selection, since that is the only case sensitive to compressed spectra where an excess has been observed. Again we only implement signal regions and not control regions. It should be noted that the on-shell selection has previously been recast in {\sc GAMBIT}, and we have also implemented that analysis in \hackanalysis. The results of the off-shell $WZ$ analysis are interpreted within the same simplified higgsino and wino/bino($\pm$) frameworks introduced for ATLAS-SUSY-2018-16.

\begin{table*}
\begin{center}\renewcommand{\arraystretch}{1.3}\setlength{\tabcolsep}{6pt}
\begin{tabular}{l c c c c}
& \multicolumn{4}{c}{Preselection requirements} \\
\cline{2-3}\cline{4-5} 
\rule{0pt}{\dimexpr.7\normalbaselineskip+1mm}
Variable                        & \SRlowzj & \SRlownj & \SRhighzj & \SRhighnj \\[0.1cm]
\midrule
\nlbl, \nlsig                   & \multicolumn{4}{c}{= 3}                    \\
$n_\mathrm{OSSF}$                          & \multicolumn{4}{c}{$\geq 1$}               \\
$m_{\ell\ell}^{\text{max}}$ [GeV]                  & \multicolumn{4}{c}{$<75$}                  \\
$m_{\ell\ell}^{\text{min}}$ [GeV]                  & \multicolumn{4}{c}{$\in [1,75]$}           \\
\nbjets                         & \multicolumn{4}{c}{$= 0$}                  \\
$\min \Delta R_{3\ell}$                       & \multicolumn{4}{c}{$>0.4$}                 \\
\midrule
Resonance veto \mllmin [GeV]\ \ \ \ \ \ \   & \multicolumn{3}{c|}{$\notin [3,3.2]$, $\notin [9,12]$} & - \\
Trigger                         & \multicolumn{2}{c|}{(multi-)lepton\hspace{5ex}\ } & \multicolumn{2}{c}{((multi-)lepton\ ||\ $E_{\text{T}}^{\text{miss}}$)} \\
$n_{\text{jets}}^{30~\text{GeV}}$                        & $= 0$ & $\geq 1$ & $= 0$ & $\geq 1$                                                  \\
$E_{\text{T}}^{\text{miss}}$ [GeV]                     & $<50$  & $<200$ & $>50$ & $>200$                                                     \\
$E_{\text{T}}^{\text{miss}}$ significance                         & $>1.5$ & $>3.0$ & $>3.0$ & $>3.0$                                                    \\
\ptl{1}, \ptl{2}, \ptl{3} [GeV] & \multicolumn{3}{c|}{$>10$} & \ $>4.5(3.0)$ \ for $e(\mu)$ \\
$|\mtl-m_Z|$ [GeV]             & \multicolumn{2}{c|}{$>20$ {($\lW=e$ only)}\hspace{5ex}\ } & \multicolumn{2}{c}{-}       \\
$\min \Delta R_{\mathrm{OSSF}}$                      & \multicolumn{2}{c|}{$[0.6,2.4]$ {($\lW=e$ only)}\hspace{5ex}\ } & \multicolumn{2}{c}{-} \\
\end{tabular}
\end{center}
\caption{\label{tab:ofs:presel}Summary of the preselection criteria applied in the SRs of the \ofs \WZ selection in ATLAS-SUSY-2019-09. In rows where only one value is given, it applies to all regions.}
\end{table*}

\begin{table*}
\begin{center}\renewcommand{\arraystretch}{1.2}\setlength{\tabcolsep}{8pt}
\begin{tabular}{l c c c c c c c c c} 
& \multicolumn{9}{c}{Selection requirements} \\
\cline{2-10} 
Variable        & \texttt{a} & \texttt{b} & \texttt{c} & \texttt{d} & \texttt{e} & \texttt{f1} & \texttt{f2} & \texttt{g1} & \texttt{g2} \\
\midrule
\mllmin [GeV]  & [1, 12] & [12, 15] & [15, 20] & [20, 30] & [30, 40] & \multicolumn{2}{c}{[40, 60]} & \multicolumn{2}{c}{[60, 75]} \\
\hline\hline 
\rule{0pt}{\dimexpr.7\normalbaselineskip+1mm}
& \multicolumn{9}{c}{\rSR{\rWZof}{\rlo\rmet}{}~common} \\[0.1cm]
\cline{2-10} 
\mllmax [GeV]     & $\times$ & $<60$ & $<60$ & $<60$ & $<60$ & - & - & - & - \\
\mtmllmin [GeV]   & $\times$ & $<50$ & $<50$ & $<50$ & $<60$ & $<60$ & $>90$ & $<60$ & $>90$ \\
\mttwoh [GeV]     & $\times$ & $<115$ & $<120$ & $<130$ & - & - & - & - & - \\
$\min \Delta R_{\text{OSSF}}$        & $\times$ & $<1.6$ & $<1.6$ & $<1.6$ & - & - & - & - & - \\
\ptl{1},\ptl{2},\ptl{3} [GeV]\ \ \ \ \ \ & $\times$ & $>10$ & $>10$ & $>10$ & $>10$ & $>15$ & $>15$ & $>15$ & $>15$ \\
\cline{2-10} 
\rule{0pt}{\dimexpr.7\normalbaselineskip+1mm}
& \multicolumn{9}{c}{\rSR{\rWZof}{\rlo\rmet}{-\rzj}} \\[0.1cm]
\cline{2-10} 
$|\tv{p}_{\text{T}}^{\text{lep}}|/E_{\text{T}}^{\text{miss}}$      & $\times$ & $<1.1$& $<1.1$  & $<1.1$ & $<1.3$ & $<1.4$ & $<1.4$ & $<1.4$ & $<1.4$ \\
\mtl [GeV] & $\times$ & - & - & - & - & $>100$ & $>100$ & $>100$ & $>100$ \\
\cline{2-10} 
\rule{0pt}{\dimexpr.7\normalbaselineskip+1mm}
& \multicolumn{9}{c}{\rSR{\rWZof}{\rlo\rmet}{-\rnj}} \\[0.1cm]
\cline{2-10} 
$|\tv{p}_{\text{T}}^{\text{lep}}|/E_{\text{T}}^{\text{miss}}$      & $\times$ & $<1.0$& $<1.0$  & $<1.0$ & $<1.0$ & $<1.2$ & $<1.2$ & $<1.2$ & $<1.2$ \\
\toprule
\rule{0pt}{\dimexpr.7\normalbaselineskip+1mm}
& \multicolumn{9}{c}{\rSR{\rWZof}{\rhi\rmet}{}~common} \\[0.1cm]
\cline{2-10} 
\mttwoh [GeV]     & $<112$ & $<115$ & $<120$ & $<130$ & $<140$ & $<160$ & $<160$ & $<175$ & $<175$ \\
\cline{2-10} 
\rule{0pt}{\dimexpr.7\normalbaselineskip+1mm}
& \multicolumn{9}{c}{\rSR{\rWZof}{\rhi\rmet}{-\rzj}} \\[0.1cm]
\cline{2-10} 
\ptl{1},\ptl{2},\ptl{3} [GeV] & $\times$ & \multicolumn{7}{c}{$>25$, $>15$, $>10$} & \\
\mtmllmin [GeV]   & $\times$ & $<50$ & $<50$ & $<60$ & $<60$ & $<70$ & $>90$ & $<70$ & $>90$ \\
\cline{2-10} 
\rule{0pt}{\dimexpr.7\normalbaselineskip+1mm}
& \multicolumn{9}{c}{\rSR{\rWZof}{\rhi\rmet}{-\rnj}} \\[0.1cm]
&   &   &   &   &   & \multicolumn{2}{c}{f} & \multicolumn{2}{c}{g} \\
\cline{2-10} 
\ptl{1},\ptl{2},\ptl{3} [GeV] & & \multicolumn{7}{c}{$>4.5\,(3.0)$ \ for $e\,(\mu)$} & \\
$|\tv{p}_{\text{T}}^{\text{lep}}|/E_{\text{T}}^{\text{miss}}$     & $<0.2$ & $<0.2$ & $<0.3$ & $<0.3$ & $<0.3$ & \multicolumn{2}{c}{$<1.0$} & \multicolumn{2}{c}{$<1.0$} \\
\end{tabular}
\end{center}
\caption{
Summary of the selection criteria for exclusive SRs for the \ofs \WZ selection in ATLAS-SUSY-2019-09. SRs are given by the combination(s) of preselection cuts in Table~\ref{tab:ofs:presel} with those listed in this table (\SRlowzj\!\texttt{a}, etc.). The symbol $\times$ indicates regions not considered.
}
\label{tab:ofs:sel1}
\end{table*}

The preselection requirements for ATLAS-SUSY-2019-09 are listed in Table~\ref{tab:ofs:presel}. Exactly three leptons are required, from which set it must be possible to form at least one pair of OSSF leptons that can (naturally) be quite soft. The analysis additionally includes requirements on the minimum and maximum invariant masses $m_{\ell\ell}^{\text{min,max}}$ that can be computed from the OSSF pair(s),\footnote{This analysis is also performed to identify the lepton most likely associated with the $W^{\pm}$ decay.} as well as isolation requirements ($\Delta R_{3\ell} = \min \Delta R(\ell_i,\ell_j)\ \forall\ \{\ell_i,\ell_j\}$ and $\Delta R_{\text{OSSF}} = \min \Delta R(\ell_i,\ell_j)\ \forall\ \text{OSSF}\ \{\ell_i,\ell_j\}$) and a $Z$-mass window veto. The analysis further involves low- and high-\MET signal regions, with the \MET threshold being either $E_{\text{T}}^{\text{miss}} = 50$~GeV or 200~GeV depending on the multiplicity of jets with transverse momentum $p_{\text{T}}(j) > 30$~GeV: the higher \MET threshold applies to the SRs with $\geq 1$ jet since the electroweakinos have harder recoil in these events. An object-based \MET \emph{significance}, derived from uncertainties on the measurement of individual physics objects \cite{ATLAS:2018uid}, is additionally used to control for missed or mismeasured visible objects.

\begin{figure*}
    \centering
    \includegraphics[width=0.47\textwidth]{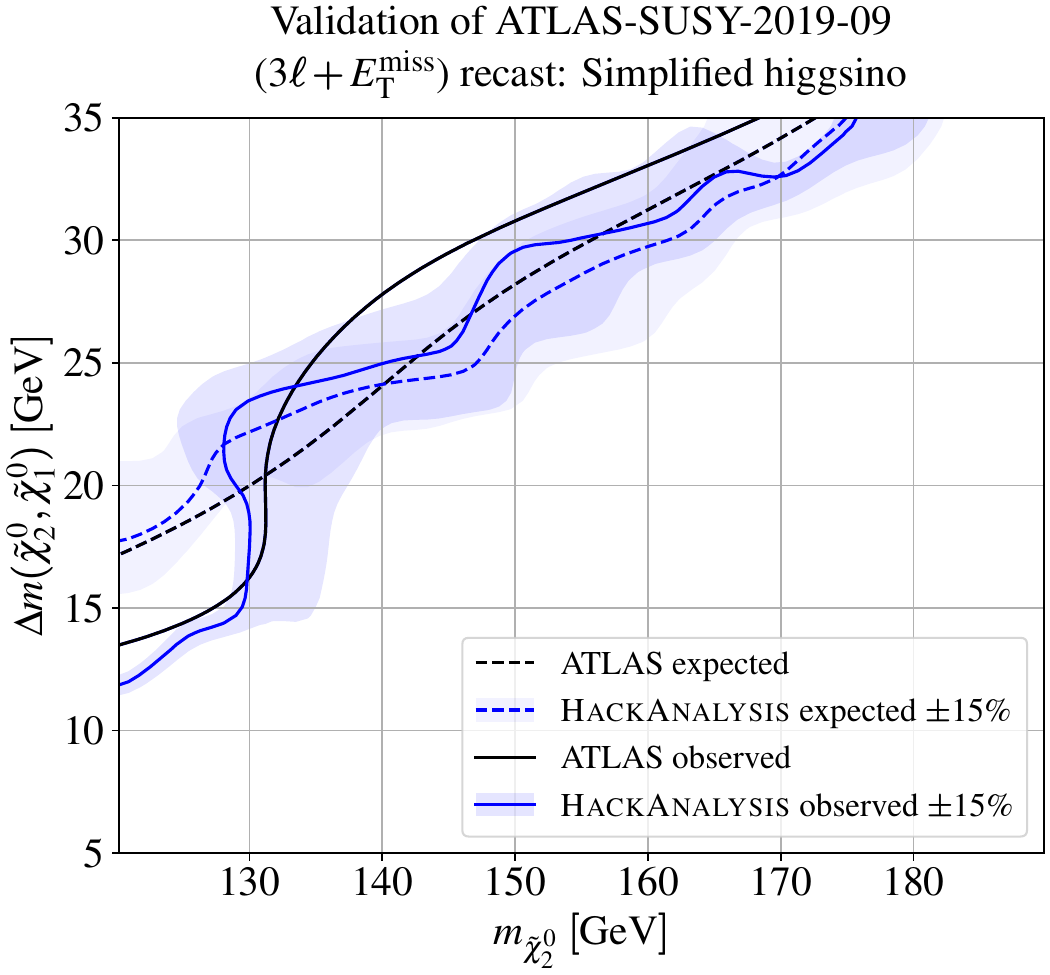}\hfill
    \includegraphics[width=0.47\textwidth]{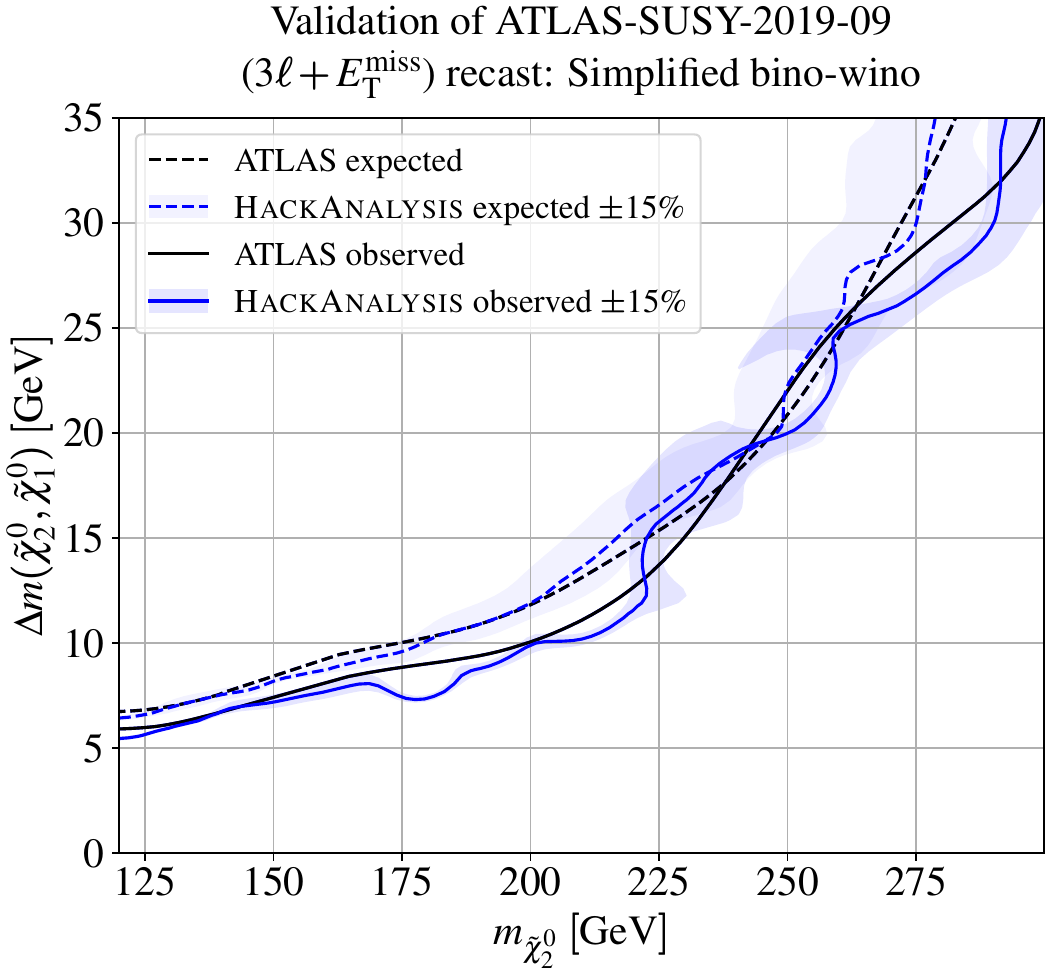}
    \caption{Comparison of \textsc{HackAnalysis} exclusions to official ATLAS results for ATLAS-SUSY-2019-09. Left panel shows exclusions in the simplified higgsino model while right panel refers to the simplified bino-wino scenario. \textsc{HackAnalysis} uncertainty bands correspond to 15\% variation of the signal strength.}\label{fig:3lValid}
\end{figure*}

The selection requirements are listed in Table~\ref{tab:ofs:sel1}, with the SR concatenations employed to conduct an inclusive analysis specified in Table~\ref{tab:ofs:disc} in \ref{a1}. A fairly large number of exclusive SRs are defined for the off-shell $WZ$ analysis alone. These SRs are principally distinguished by non-overlapping requirements on $m_{\ell\ell}^{\text{min}}$, since this observable has an endpoint at $\Delta m(\tilde{\chi}^0_2,\tilde{\chi}^0_1)$ for the targeted SUSY signals. Additional cuts are imposed on lepton $p_{\text{T}}$ and isolation, as well as on the transverse mass $m_{\text{T}}^{\text{mllmin}}$ of the lepton assigned to the $W$ boson and the stransverse mass~\cite{Lester:1999tx, Cheng:2008hk} $m_{\text{T}2}^{100}$ of the $(2+1)$-lepton system (\ie, the combined system comprising the OSSF pair associated with the $Z$ and the lepton assigned to the $W^{\pm}$) with a test mass of $m_{\chi} = 100$~GeV. There are finally cuts on the trilepton invariant mass $m_{3\ell}$ and the ratio $|\tv{p}_{\text{T}}^{\text{miss}}|/E_{\text{T}}^{\text{miss}}$ of the trilepton transverse momentum to the missing transverse energy, the latter of which targets events with \MET resulting from massive $\tilde{\chi}^0_1$ escaping detection.  The ATLAS Collaboration reports expected and observed yields in the 31 exclusive signal regions (which we do not reproduce here for want of space) and in the 17 inclusive signal regions constructed according to Table~\ref{tab:ofs:disc}. The inclusive yields are shown in our Table~\ref{tab:results:offShell_discSRs}, in \ref{a1}, along with the discovery $p$-value computed by ATLAS for each signal region.

As stated above, the on-shell and off-shell selections (but not the $Wh$ selection) have now been implemented in \hackanalysis; here we discuss only the recast and validation of the off-shell version, since that is the one used in the present work. For this purpose, the ATLAS Collaboration provided a pseudocode, efficiencies for the reconstruction of the leptons in the form of plots (which could be scraped), cutflows, and a \pyhf background model. Yet the particulars of the off-shell selection nevertheless present an interesting challenge to recast: for instance, only one class of signal regions contains a straightforward cut on minimum missing transverse energy of $200$~GeV; as mentioned above, the others either require $E_{\text{T}}^{\text{miss}} > 50$ GeV or in fact have \emph{upper} bounds of 50 or 200~GeV. The additional multilepton triggers in those signal regions compensate for the known inefficiency of the \MET trigger below 200~GeV. In addition, the object-based \MET significance, which is not entirely trivial to implement in a recast, is used only in the off-shell selection. This is perhaps why only the on-shell $WZ$ analysis has been implemented in {\sc GAMBIT}, since it does not rely on this quantity. 

In order to model the lepton trigger efficiencies, data from dedicated ATLAS studies \cite{ATLAS:2019dpa,ATLAS:2020gty} were used; the trigger efficiencies for events featuring dielectrons and single, double and triple muons can be found, and from these the probability that a given event will activate one of the triggers can be inferred by checking each case in turn. For the \MET significance, defined and studied in \cite{ATLAS:2018uid}, it is possible to compute an uncertainty associated to each physics object (jets and leptons in this case, but in principle also photons) using the data in that reference. The data there give the uncertainty on the resolution of the momenta for any such object along its direction of motion, $\sigma_{\parallel},$ and perpendicular to it, $\sigma_\perp,$ in the plane perpendicular to the beam axis. The aim is to determine a covariance matrix of uncertainties, given by
\begin{align}
\mathbf{V}_{xy} &\equiv \begin{pmatrix}\sigma_{xx}^2 & \sigma_{xy}^2 \\ \sigma_{xy}^2 & \sigma_{yy}^2 \end{pmatrix} = R^{-1} (\phi) \begin{pmatrix} \sigma_{\perp}^2 & 0 \\ 0 & \sigma_{\parallel}^2 \end{pmatrix} R(\phi),
\end{align}
introducing a rotation matrix $R(\phi)$. The sum of all of these matrices for each object can then be added to a component from the ``soft term,'' which we take to be diagonal with magnitude $(8.9~\mathrm{GeV})^2.$ The \MET significance $\mathcal{S}(E_{\text{T}}^{\text{miss}})$ is then given as
\begin{align}
   [\mathcal{S}(E_{\text{T}}^{\text{miss}})]^2 \equiv \mathbf{p}_\perp^{\rm miss} \cdot (\mathbf{V}_{xy}^{\rm total} )^{-1} \cdot \mathbf{p}_\perp^{\rm miss}.
\end{align}
In \hackanalysis we evaluate this in the analysis from the relevant reconstructed objects. We use uncertainties on each class of object scraped from data in \cite{ATLAS:2018uid} and parameterized as a function of transverse momentum.

To validate our recast of the off-shell $WZ$ selection, we again compare with both the simplified higgsino and wino/bino(+) scenarios using the same simulated events as for the $2\ell$ analysis, plus events from additional parameter points where necessary (while the neutralino decays are biased, the chargino decays are not forced to be hadronic for $\tilde{\chi}^0_2 \tilde{\chi}^{\pm}_1$ events in order to make this possible). We simulate 239 simplified higgsino and 475 simplified bino-wino points for this validation. The comparison of exclusion plots for these scenarios is shown in Figure~\ref{fig:3lValid}. The signals provided for ATLAS-SUSY-2019-09 do not carry a (non-negligible) statistical uncertainty like those in ATLAS-SUSY-2018-16; the only uncertainty estimate in Figure~\ref{fig:3lValid} therefore comes from the 15\% variation of our own signal strengths. In both model scenarios we find very good agreement with the official results.

\subsection{Monojet searches: ATLAS-EXOT-2018-06 and CMS-EXO-20-004}

\begin{figure}
    \centering
    \includegraphics[scale=1]{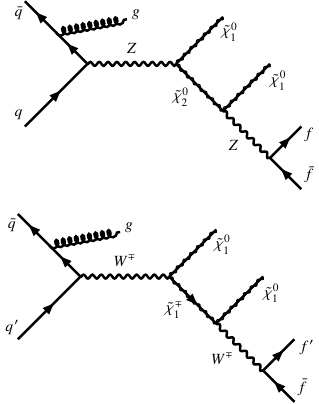}
    \caption{\label{fig:monojetGraph}Representative diagrams for monojet signals originating from light electroweakino pair production. The fermions $f,f'$ must be soft.}
\end{figure}

\renewcommand{\arraystretch}{1.1}\setlength{\tabcolsep}{6pt}
\begin{table}
\centering
\begin{tabular}{l c@{\hspace{3ex}}c@{\hspace{3ex}}r}
$E_{\text{T}}^{\rm miss}$ [GeV] & Background & Observed & $\chi$ [$\sigma$] \\
\midrule
$[200,250]$  & \ \ $1783000 \pm 26000$\ \ & $1791624$ & $0.3$ \\
$[250,300]$  & $753000 \pm 9000$ & $752328$ & $-0.1$ \\
$[300,350]$  & $314000 \pm 3500$ & $313912$ & $-0.0$ \\
$[350,400]$  & $140100 \pm 1600$ & $141036$ & $0.6$ \\
$[400,500]$  & $101600 \pm 1200$ & $102888$ & $1.1$ \\
$[500,600]$  & $29200 \pm 400$ & $29458$ & $0.6$ \\
$[600,700]$  & $10000 \pm 180$ & $10203$ & $1.1$ \\
$[700,800]$  & $3870 \pm 80$ & $3986$ & $1.4$ \\
$[800,900]$  & $1640 \pm 40$ & $1663$ & $0.6$ \\
$[900,1000]$  & $754 \pm 20$ & $738$ & $-0.8$ \\
{\color{red}$[1000,1100]$}  & {\color{red}$359 \pm 10$} & {\color{red}$413$} & {\color{red}$5.4$} \\
$[1100,1200]$  & $182 \pm 6$ & $187$ & $0.8$ \\
$[1200,\infty]$  & $218 \pm 9$ & $207$ & $-1.2$ \\
\end{tabular}
\caption{\label{tab:atlasmono}Number of expected and observed events by signal bin for the ATLAS monojet search. For each bin we give the $\chi$-value defined as (observed$-$background)/(background uncertainty). }
\end{table}

The monojet searches by ATLAS \cite{ATLAS:2021kxv} and CMS \cite{CMS:2021far} were previously recast in \madanalysis \cite{DVN/IRF7ZL_2021, DVN/REPAMM_2023}; for statistical information, single-bin data for the ATLAS search must be used (since no statistical information was provided) while CMS provided simplified-likelihood yields and a covariance matrix.  These searches are interpreted within different model frameworks but can be reinterpreted for electroweakino pair production followed by decays to the LSP and soft Standard Model fermions, as exemplified in Figure~\ref{fig:monojetGraph}. In principle, these analyses' rejection of leptons and (additional) jets with $p_{\text{T}}$ of more than a few GeV makes them sensitive to very small electroweakino mass splittings and thus complementary to the soft-lepton searches discussed above. In previous work \cite{Agin:2023yoq}, however, we discussed the presence of excesses in both of those searches; in the ATLAS analysis these occur for $E_{\text{T}}^{\text{miss}} > 400$ GeV, with a $5\sigma$ local excess in the bin with $E_{\text{T}}^{\text{miss}} \in [1000,1100]$ GeV. In an attempt to compare the ATLAS and CMS excesses, we show the equivalent information for the two analyses in Tables~\ref{tab:atlasmono} and \ref{tab:cmsmono}. Intriguingly, the CMS analysis shows many excesses in the 2018 dataset with apparent local significances approaching five standard deviations; the CMS bin closest to the ATLAS excess bin, with $E_{\text{T}}^{\text{miss}} \in [1020,1090]$~GeV, has a local excess of $3.88\sigma$. But the 2016 dataset also contains excesses of up to $3\sigma$ in bins with $E_{\text{T}}^{\text{miss}}$ ranging from 550~GeV to 740~GeV. Altogether, while it is relatively easy to infer the existence of some excess from both tables, it is not so straightforward to map the ATLAS excess onto the CMS analysis. Further investigation is probably warranted.

A further comment about the CMS monojet excesses is also in order. In Figure 4 of \cite{CMS:2021far}, plots are presented showing the number of events observed and the predicted number of background events prior to and post fitting to all the background and signal regions, and while mild excesses can be seen, they do not correspond to the values that are given in Table~\ref{tab:cmsmono}; this can cause some confusion. The values presented here, which we took from the {\sc HEPData} entry, correspond to values computed by CMS to be used as inputs for the simplified likelihood computation, which had a different fitting procedure involving just the control regions, as it should be when we have integrated them out. It is not surprising that the excesses are diminished after a fit to all regions, but perhaps more surprising is that the pre-fit results shown by CMS have a different pattern of deviations: the difference comes from the behavior of the control regions. 

For the purposes of this work, the recasts have been adapted into \hackanalysis so that the same signal events could be processed through all four analyses at the same time (where appropriate; \ie, when no bias is imposed on the signal generation); this also obviates the need for a {\sc Delphes}~\cite{deFavereau:2013fsa} simulation. The validation for CMS-EXO-20-004 was performed using the same models and cutflows as the \madanalysis validation~\cite{CMS:2021far}, obtaining agreement on the cutflows up to the final cut within within $3\%$, and within $10\%$ or $2\sigma$ statistical uncertainty on all missing-energy bins. Validation material for these recasts will be provided elsewhere.

%% file: Sections/3_MSSM.tex
\section{Overlapping excesses in a realistic MSSM with bino-wino LSP}
\label{s3}

In previous work \cite{Agin:2023yoq}, we demonstrated that the mild excesses in the soft-dilepton and monojet channels overlap to some degree for scenarios featuring compressed ``pure'' higgsinos. Since one aim of this work is to understand if the higgsino scenario is optimal or even unique with respect to these excesses, we use the recasts described in the previous section to reinterpret the analyses for a suite of models. We first consider a limit of the MSSM in which the higgsinos are decoupled, which is not only orthogonal to the light-higgsino scenario (both simplified and realistic) but also distinct from the bino-wino scenario considered in ATLAS-SUSY-2018-16.

\begin{table*}\renewcommand{\arraystretch}{1.3}
\resizebox{\textwidth}{!}{%
\begin{tabular}{l c@{\hspace{3ex}}c@{\hspace{3ex}}r@{\hspace{1.5ex}}| c@{\hspace{3ex}}c@{\hspace{3ex}}r@{\hspace{1.5ex}}|c@{\hspace{3ex}}c@{\hspace{3ex}}r}
\multirow{2}{*}[-0.4ex]{$E_{\text{T}}^{\rm miss}$ [GeV]} & \multicolumn{3}{c|}{2016} &  \multicolumn{3}{c|}{2017} &  \multicolumn{3}{c}{2018} \\
\rule{0pt}{3ex}& Background & Observed & $\chi$ [$\sigma$] & Background & Observed & $\chi$ [$\sigma$] & Background & \ \ \ Observed\ \ \ & \ \ \ $\chi$ [$\sigma$] \\ \midrule
$[250,280]$  & $135004.86 \pm 3514.65$ & $136865$ & $0.5$ & $169737.77 \pm 3706.95$ & $176171$ & $1.74$ & $178470.12 \pm 3615.00$ & $191829$ &$ 3.70$ \\
$[280,310]$  & $73681.17 \pm 1904.73$ & $74340$ & $0.3$ & $93266.64 \pm 1975.68$ & $96067$ & $1.42$ & $97428.42 \pm 1896.81$ & $104522$ &$ 3.74$ \\
$[310,340]$  & $42448.18 \pm 822.14$ & $42540$ & $0.1$ & $54175.54 \pm 1117.22$ & $54789$ & $0.55$ & $55592.19 \pm 1059.45$ & $59398$ &$ 3.59$ \\
$[340,370]$  & $25541.24 \pm 513.41$ & $25316$ & $-0.4$ & $32232.17 \pm 665.78$ & $32767$ & $0.80$ & $33037.13 \pm 618.12$ & $35833$ &$ 4.52$ \\
$[370,400]$  & $15454.11 \pm 315.39$ & $15653$ & $0.6$ & $19798.25 \pm 413.29$ & $20209$ & $0.99$ & $20714.72 \pm 391.42$ & $21854$ &$ 2.91$ \\
$[400,430]$  & $10162.68 \pm 195.49$ & $10092$ & $-0.4$ & $12816.66 \pm 266.21$ & $12910$ & $0.35$ & $13209.01 \pm 250.28$ & $14266$ &$ 4.22$ \\
$[430,470]$  & $8473.07 \pm 168.65$ & $8298$ & $-1.0$ & $10626.92 \pm 217.70$ & $10653$ & $0.12$ & $10991.22 \pm 213.74$ & $11730$ &$ 3.46$ \\
$[470,510]$  & $4853.26 \pm 106.33$ & $4906$ & $0.5$ & $6407.87 \pm 133.23$ & $6487$ & $0.59$ & $7836.98 \pm 152.03$ & $8639$ &$ 5.28$ \\
$[510,550]$  & $2960.10 \pm 67.10$ & $2987$ & $0.4$ & $3903.66 \pm 84.55$ & $3955$ & $0.61$ & $4914.64 \pm 101.04$ & $5406$ &$ 4.86$ \\
$[550,590]$  & $1906.69 \pm 50.32$ & $2032$ & $2.5$ & $2519.86 \pm 58.43$ & $2587$ & $1.15$ & $3065.69 \pm 66.79$ & $3285$ &$ 3.28$ \\
$[590,640]$  & $1498.36 \pm 39.65$ & $1514$ & $0.4$ & $1929.79 \pm 45.84$ & $1957$ & $0.59$ & $2428.39 \pm 55.08$ & $2671$ &$ 4.40$ \\
$[640,690]$  & $839.22 \pm 26.10$ & $926$ & $3.3$ & $1181.81 \pm 30.83$ & $1151$ & $-1.00$ & $1441.83 \pm 35.64$ & $1613$ &$ 4.80$ \\
$[690,740]$  & $523.39 \pm 17.78$ & $557$ & $1.9$ & $713.79 \pm 21.29$ & $721$ & $0.34$ & $903.52 \pm 24.56$ & $965$ &$ 2.50$ \\
$[740,790]$  & $322.84 \pm 13.17$ & $316$ & $-0.5$ & $456.23 \pm 16.28$ & $455$ & $-0.08$ & $603.29 \pm 18.99$ & $640$ &$ 1.93$ \\
$[790,840]$  & $221.07 \pm 10.19$ & $233$ & $1.2$ & $294.63 \pm 11.86$ & $298$ & $0.28$ & $385.55 \pm 14.01$ & $424$ &$ 2.74$ \\
$[840,900]$  & $167.22 \pm 9.56$ & $172$ & $0.5$ & $231.17 \pm 10.18$ & $244$ & $1.26$ & $271.75 \pm 10.78$ & $319$ &$ 4.38$ \\
$[900,960]$  & $106.22 \pm 6.98$ & $101$ & $-0.7$ & $146.41 \pm 7.44$ & $146$ & $-0.05$ & $176.46 \pm 8.49$ & $191$ &$ 1.71$ \\
$[960,1020]$  & $87.38 \pm 6.92$ & $65$ & $-3.2$ & $87.26 \pm 5.39$ & $89$ & $0.32$ & $117.78 \pm 6.52$ & $117$ &$ -0.12$ \\
$[1020,1090]$  & $52.37 \pm 4.72$ & $46$ & $-1.4$ & $69.26 \pm 4.73$ & $65$ & $-0.90$ & $75.92 \pm 5.18$ & $96$ &$ 3.88$ \\
$[1090,1160]$  & $24.70 \pm 3.15$ & $26$ & $0.4$ & $32.27 \pm 3.23$ & $39$ & $2.08$ & $48.50 \pm 3.76$ & $55$ &$ 1.73$ \\
$[1160,1250]$  & $25.23 \pm 3.05$ & $31$ & $1.9$ & $30.65 \pm 3.12$ & $35$ & $1.39$ & $35.88 \pm 3.39$ & $43$ &$ 2.10$ \\
$[1250,\infty]$  & $26.61 \pm 3.10$ & $29$ & $0.8$ & $37.35 \pm 3.33$ & $40$ & $0.80$ & $53.36 \pm 4.06$ & $70$ &$ 4.10$ \\
\end{tabular}%
}
\caption{\label{tab:cmsmono}Number of expected and observed events by signal bin and year for the CMS monojet search. For each bin we give the $\chi$-value defined as (observed$-$background)/(background uncertainty). }
\end{table*}
\renewcommand{\arraystretch}{1.0}

One reason to explore the case of a bino-like LSP is the suitability of the bino as a light (sub-TeV) dark matter candidate: unlike higgsinos, which annihilate too quickly and yield underabundant dark matter unless the LSP attains roughly 1~TeV in mass, scenarios with heavy higgsinos can predict much slower electroweakino (co)annihilation. This is certainly one motivation behind ATLAS' choice to target bino-wino (N)LSP scenarios in their soft-lepton analyses. On the other hand, ATLAS makes some artificial choices in the generation of the parameter cards for the simplified bino-wino signals in the soft-lepton analyses (we elaborate on this below). It is therefore worthwhile, and complementary to our work in~\cite{Agin:2023yoq}, to define a ``realistic'' bino-wino parameter space and interpret the various analyses within this framework. Our scenario most closely resembles the ATLAS wino/bino(+) model but differs in a few subtle but important ways; we discuss these differences further below using a pair of benchmark points.

We are interested in a scenario where the light electroweakinos are composed entirely of gauginos; \ie, the higgsinos are decoupled. In the limit $M_1 < M_2 \ll \mu$, which guarantees a bino-like LSP, the light electroweakino masses are given approximately by \cite{PhysRevD.37.2515}
\begin{align}
\nonumber m_{\tilde{\chi}^0_1} &\simeq M_1 + (m_Z \sin \theta_{\text{w}})^2\, \frac{M_1 + \mu \sin 2\beta}{M_1^2 - \mu^2},\\
\nonumber m_{\tilde{\chi}^{\pm}_1} &\simeq M_2 + m_W^2\,\frac{M_2 + \mu \sin 2\beta}{M_2^2 - \mu^2},\\
\text{and}\ \ \ m_{\tilde{\chi}^0_2} &\simeq M_2 + (m_Z \cos \theta_{\text{w}})^2\, \frac{M_2 + \mu \sin 2\beta}{M_2^2 - \mu^2}
\end{align}
with $\theta_{\text{w}}$ denoting the weak mixing angle, $m_Z$ and $m_W$ the masses of the $Z$ and $W$ bosons, and $\tan \beta = v_u v_d^{-1}$ the ratio of the Higgs vacuum expectation values. For NLSP-LSP mass splittings $\Delta m(\tilde{\chi}^0_2,\tilde{\chi}^0_1)$ in the vicinity of 20~GeV, the bino-like $\tilde{\chi}^0_1$ is a viable dark matter candidate with a relic abundance close to the value $\Omega h^2 = 0.12 \pm 0.001$ reported by the Planck Collaboration assuming a standard cosmological history~\cite{Planck:2018vyg}. In this parameter space, the correct relic abundance is achieved principally via (co)annihilations of the other light electroweakinos, notably $\tilde{\chi}^0_2 \tilde{\chi}^-_1 \to \bar{u}d$, $\tilde{\chi}^0_2 \tilde{\chi}^0_2 \to W^+ W^-$, and $\tilde{\chi}^+_1 \tilde{\chi}^-_1 \to W^+ W^-,ZZ$.

\subsection{Parameter space scan and event generation}
\label{s3.1}

\begin{table*}\renewcommand{\arraystretch}{1.3}
\centering
    \begin{tabular}{l l l}
 & ATLAS wino/bino(+) & Decoupled higgsinos \\
\midrule
$\{m_{\tilde{\chi}^0_1},m_{\tilde{\chi}^0_2},m_{\tilde{\chi}^{\pm}_1}\}$~[GeV]\ \ \ \ \ \ & $\{235,250,250\}$ & $\{238.9,255.6,255.8\}$ \\[0.04cm]
$(N_{11}^2,N_{12}^2,N_{13}^2+N_{14}^2)$  & $(0.9730,0.0028,0.0243)$\ \ \ \ \ \ & $(0.9991,0.0004,0.0005)$ \\[0.04cm]
$(N_{21}^2,N_{22}^2,N_{23}^2+N_{24}^2)$ & $(0.0099,0.8928,0.0972)$ & $(0.0004,0.9980,0.0016)$ \\[0.04cm]
\midrule
\multirow{5}{*}[-1ex]{\makecell{$\tilde{\chi}^0_2$ decay modes\\(branching fraction)}} & $\tilde{\chi}^0_1 Z^*$ (100\%)  & $\tilde{\chi}^0_1 q\bar{q}$ (72.8\%) \\[0.04cm]
& & $\tilde{\chi}^0_1 \nu \bar{\nu}$ (15.9\%) \\[0.04cm]
& & $\tilde{\chi}^0_1 e^+ e^-$ (3.98\%) \\[0.04cm]
& & $\tilde{\chi}^0_1 \mu^+ \mu^-$ (3.98\%) \\[0.04cm]
& & $\tilde{\chi}^0_1 \tau^+ \tau^-$ (3.27\%) \\[0.04cm]
\midrule
\multirow{4}{*}[-0.5ex]{\makecell{$\tilde{\chi}^-_1$ decay modes\\(branching fraction)}} & $\tilde{\chi}^0_1 W^{-,*}$ (100\%) & $\tilde{\chi}^0_1 \bar{u}d$ (66.8\%)   \\[0.04cm]
& & $\tilde{\chi}^0_1 e^-\bar{\nu}_e$ (11.3\%) \\[0.04cm]
& & $\tilde{\chi}^0_1 \mu^-\bar{\nu}_{\mu}$ (11.3\%) \\[0.04cm]
& & $\tilde{\chi}^0_1 \tau^- \bar{\nu}_{\tau}$ (10.6\%) \\[0.04cm]
\end{tabular}
\caption{Comparison of neutralino mixings and decay modes in the ATLAS simplified bino-wino scenario [wino/bino(+)] and our decoupled-higgsinos scenario with bino-like LSP and wino-like NLSP.}\label{tab:MSSMbenchmarkComp}
\end{table*}

To explore this parameter space and produce the various inputs needed for event generation, we use \textsc{BSMArt} to steer a random scan of the bino-wino plane $(M_1,M_2)$. The variable inputs for this scan are simply
\begin{align}
    M_1,M_2 \in [125,325]~\text{GeV},
\end{align}
and the other relevant parameters are fixed as follows:
\begin{align}
\mu &= 2~\text{TeV}\ \text{and}\ \tan \beta = 25,\\
\nonumber    M_A^2 &= 2.5 \times 10^7~\text{GeV}^2,\\
\nonumber   M_{\tilde{q}}^2 = M_{\tilde{u},\tilde{d}}^2 &= (8.25 \times 10^8,8.25 \times 10^8,1.25 \times 10^7)~\text{GeV}^2,\\
\nonumber    A_t &= 3.5~\text{TeV},\\
\nonumber   M_{\tilde{\ell}}^2 = M_{\tilde{e}}^2 &= (4.25 \times 10^8, 4.25 \times 10^8,4.25 \times 10^6)~\text{GeV}^2.
\end{align}
Here $M_{\tilde{f}}^2$ represent the non-negligible flavor-diagonal elements of the softly SUSY-breaking scalar mass matrices of the sfermions $\tilde{f}$, and $A_t$ stands for the SUSY-breaking trilinear top squark coupling. We emphasize the significant decoupling of the first- and second-generation sfermions. The fixed values guarantee decoupled higgsinos (and other BSM particles) and a SM-like Higgs, but they are not unique in this regard. This scan populates an area of the $(m_{\tilde{\chi}^0_2},\Delta m(\tilde{\chi}^0_2,\tilde{\chi}^0_1))$ plane roughly corresponding to
\begin{align}
    m_{\tilde{\chi}^0_2} \in [190,300],\ \ \ \Delta m(\tilde{\chi}^0_2,\tilde{\chi}^0_1) \in (0,27]~\text{GeV},
\end{align}
which turns out to be ideal for both the soft-lepton analyses and the dark matter (DM) relic abundance. The calculation of the MSSM mass spectrum, including electroweakino mixing matrices and decays, is handled using \textsc{SARAH} \cite{Staub:2008uz,Staub:2013tta,Goodsell:2014bna,Goodsell:2017pdq}, whose Fortran output code uses routines from the \textsc{SPheno}~\cite{Porod:2003um,Porod:2011nf} library. We use a low-scale input card. Fermion masses are computed including loop corrections, while decays are handled at leading order. Higgs masses are computed including two-loop corrections \cite{Goodsell:2014pla,Goodsell:2015ira,Braathen:2017izn}. \textsc{BSMArt} keeps benchmark points in which the lightest CP-even scalar has mass $m_h \in [122,128]$~GeV and the lightest neutralino is at least 50\% bino, which is enforced by imposing the requirement $N_{11}^2 \geq 0.5$ on the bino-$\tilde{\chi}^0_1$ element of the neutralino mixing matrix. This requirement usually, but not always, picks out $M_1 < M_2$: the smallest $\Delta m(\tilde{\chi}^0_2,\tilde{\chi}^0_1)$ can be achieved for a bino-like LSP with $M_1$ a bit heavier than $M_2$. Benchmark points with valid spectra are forwarded by \textsc{BSMArt} to \textsc{MicrOMEGAs}~\cite{Belanger:2010pz, Belanger:2013oya, Alguero:2023zol} for the calculation of the DM relic abundance and an estimate of direct-detection limits, which is performed at leading order using the \textsc{CalcHEP}~\cite{Belyaev:2012qa} model files produced by \textsc{SARAH} for the MSSM. We discard points that are ruled out by direct searches, but in order to cover a suitable range of $\Delta m(\tilde{\chi}^0_2,\tilde{\chi}^0_1)$ we retain a number of points with both over- and underabundant dark matter. For each of these points, two \texttt{SLHA} cards are produced: one, with a real neutralino mixing matrix, is used as input for \madgraph along with a UFO model produced by \textsc{SARAH}; the other, with positive electroweakino mass eigenvalues, is used as input for \textsc{Resummino}, which we use to compute the needed cross sections at next-to-next-to-leading order accuracy, including the resummation of QCD radiation at the next-to-next-to-leading (threshold) logarithmic accuracy (NNLO + NNLL). Finally, for all points surviving all prior constraints and the soft-lepton and monojet searches, we use \textsc{HiggsTools}~\cite{Bahl:2022igd} and \textsc{SModelS}~\cite{Waltenberger:2016vxp, Alguero:2021dig, MahdiAltakach:2023bdn} to perform a fast check for other experimental constraints. Points are kept only if the $p$-value inferred by \textsc{BSMArt} from the \textsc{HiggsTools} $\chi^2$ calculation exceeds $p = 0.05$ (\ie, there is no statistically significant difference between the signal model predictions and the experimental inputs used by \textsc{HiggsTools}) and the \textsc{SModelS} $r$-value is below $r = 1$.

The subtle differences between our bino-wino scenario and the ATLAS simplified wino/bino(+) scheme are shown in Table~\ref{tab:MSSMbenchmarkComp}. There we record the physical light electroweakino masses and the light neutralinos' gaugino and higgsino content, defined in terms of the elements of the neutralino mixing matrix; as well as the $\tilde{\chi}^0_2$ and $\tilde{\chi}^-_1$ branching fractions passed to \madgraph/\pythia. The left column reproduces this information from one of the wino/bino(+) SLHA cards used by the ATLAS Collaboration as input for \madgraph and made available on the \textsc{HEPData} repository for ATLAS-SUSY-2019-09.\footnote{The available points can be found under ``Resources'' at \url{https://www.hepdata.net/record/ins1866951}.} The right column reflects one of the points generated by our random scan with a mass spectrum similar to that of the ATLAS point. Inspection of all of the wino/bino(+) ATLAS cards reveals that ATLAS kept the mixing matrices and decay tables fixed and changed the masses by hand, which is artificial though acceptable for a simplified interpretation. As noted above, we instead compute the mixing matrices anew for each point. Also of note is the $\mathcal{O}(1\text{--}10)\%$ higgsino content of the ATLAS neutralinos, as opposed to our stricter decoupled-higgsino scenario. Finally, we use the branching ratios for the three-body electroweakino decays computed by the \textsc{SARAH}-\textsc{SPheno} workflow at leading order; ATLAS uses values either tabulated online\footnote{Found, for example, at \url{https://twiki.cern.ch/twiki/bin/view/LHCPhysics/SUSYCrossSections13TeVhinosplit}.} or computed via other tools, but these are not specified in the provided cards. By comparing to the online tabulated values we find that our branching ratios differ only by negligible amounts. As an aside, we note that the example parameter cards specify for each electroweakino a two-body decay to the appropriate (off-shell) weak boson with unit branching fraction, with in addition at least one \emph{allowed} three-body decay (\eg, $\tilde{\chi}^0_2 \to \tilde{\chi}^0_1 e^+ e^-$) with \emph{vanishing} branching fraction. This is the format used to force \pythia to compute the branching fractions for the electroweakinos (although it cannot compute the matrix elements and therefore simulates the decays with a flat phase space). While such cards would be acceptable for the monojet searches (and, indeed, we use the \pythia trick for those samples), the decay tables are ignored if using \textsc{MadSpin} or our workflow in any case.

\begin{figure*}
\centering
\includegraphics[width=.7\textwidth]{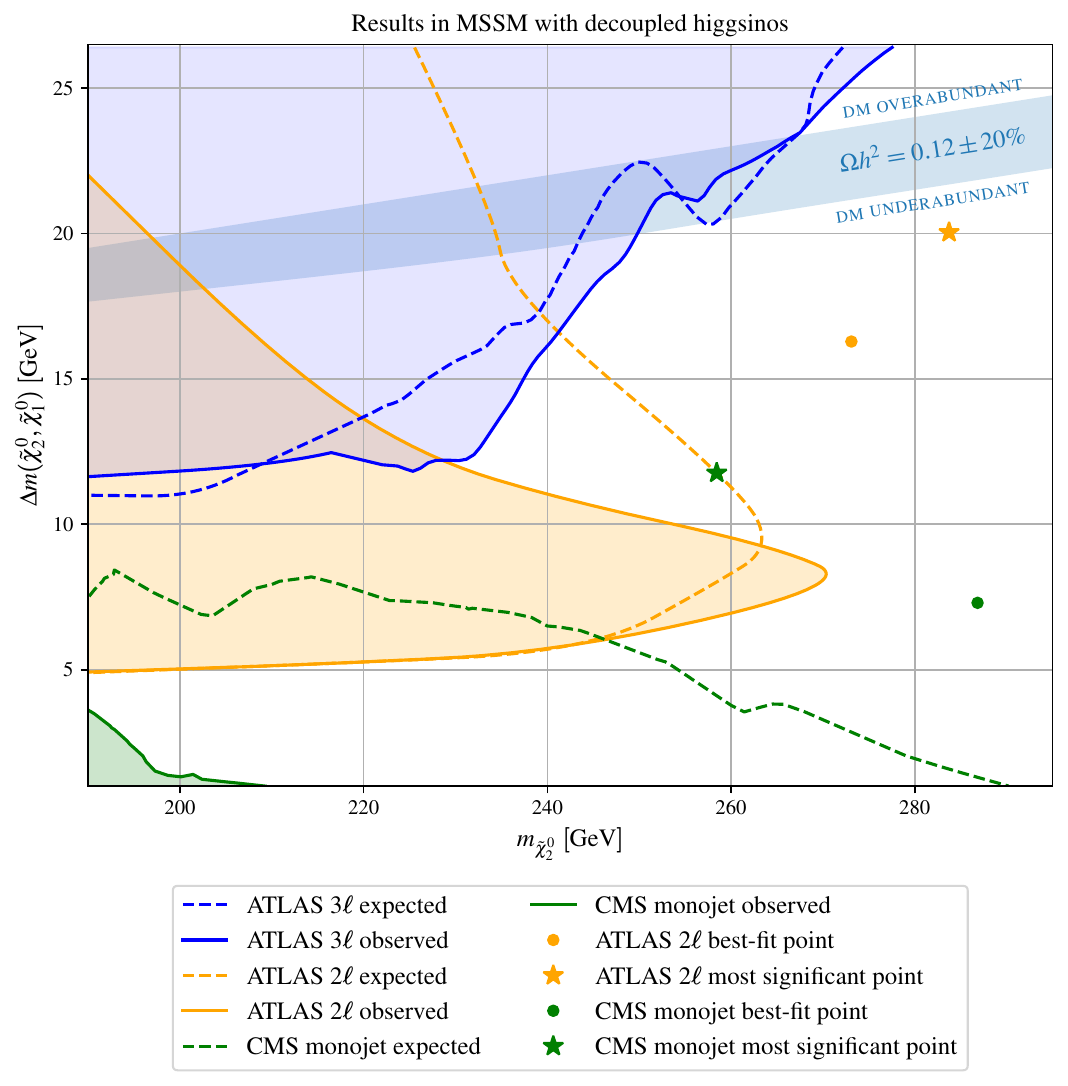}
\caption{View of the bino-wino MSSM parameter space relevant to the LHC excesses. Orange contours show limits from ATLAS-SUSY-2018-16 ($2\ell$), while blue contours refer to ATLAS-SUS-2019-09 ($3\ell$). Dashed contours show expected limits and solid contours denote the edge of the observed excluded regions, shaded correspondingly. The light blue band approximately indicates the parameter space region with relic abundance $\Omega h^2$ within 20\% of the \emph{Planck} measurement. Two interesting non-excluded points are highlighted for each of the $2\ell$ and monojet analyses.}\label{fig:mssmBinowinoPlot}
\end{figure*}

\renewcommand{\arraystretch}{1.1}
\begin{table*}\setlength{\tabcolsep}{12pt}
\centering
\resizebox{0.98\textwidth}{!}{%
    \begin{tabular}{l l@{\hspace{3ex}}l@{\hspace{3ex}}l@{\hspace{3ex}}l}
Point & {\color{mpl-orange}\scalebox{1.2}{$\bullet$}} ATLAS $2\ell$ best fit & {\color{mpl-orange}$\bigstar$} ATLAS $2\ell$ most significant & {\color{mpl-green}\scalebox{1.2}{$\bullet$}} CMS monojet best fit & {\color{mpl-green}$\bigstar$} CMS monojet most significant \\
$(m_{\tilde{\chi}^0_2},\Delta m)$ [GeV]\ \ \ \ \ & (273,\,16.3) & (284,\,20.0) & (287,\,7.30) & (258,\,11.8) \\
\midrule
$(p,\hat{\mu})$ [ATLAS $2\ell$] & (0.047,\,1.12) & (0.041,\,1.26) & $(>0.5,\,<0.1)$ & (0.290,\,0.30) \\
$(p,\hat{\mu})$ [ATLAS $3\ell$] & $(>0.5,\,<0.1)$ & (0.426,\,$<0.1$) & $(>0.5,\,<0.1)$ & $(>0.5,\,<0.1)$\\
$(p,\hat{\mu})$ [CMS monojet] & (0.098,\,1.58) & (0.065,\,2.33) & (0.049,\,1.15) & (0.044,\,1.40) \\
$(p,\hat{\mu})$ [ATLAS monojet] & (0.277,\,1.21) & (0.163,\,2.44) & (0.127,\,1.53) & (0.277,\,0.879) \\
\midrule
$M_1$ [GeV] & 248.0 & 254.6 & 269.9 & 238.2 \\
$M_2$ [GeV] & 241.7 & 251.3 & 254.0 & 228.6 \\
$m_h$ [GeV] & 127.0 & 126.5 & 126.8 & 126.8 \\
$m_{\tilde{\chi}^0_1}$ [GeV] & 256.8 & 263.7 & 279.5 & 246.7 \\
$m_{\tilde{\chi}^0_2}$ [GeV] & 273.1 & 283.7 & 286.8 & 258.4 \\
$m_{\tilde{\chi}^{\pm}_1}$ [GeV] & 273.3 & 283.9 & 287.0 & 258.6 \\
$(N_{11},N_{12},N_{13},N_{14})$ & \makecell[l]{$(0.9995,-0.0211,$\\ \hspace{5ex} $0.0232,-0.0038)$} & \makecell[l]{$(0.9996,-0.0175,$\\ \hspace{5ex} $0.0231,-0.0039)$} & \makecell[l]{$(0.9984,-0.0501,$\\ \hspace{5ex} $0.02443,-0.0043)$} & \makecell[l]{$(0.9993,-0.0284,$\\ \hspace{5ex} $0.0235,-0.0038)$} \\
$(N_{21},N_{22},N_{23},N_{24})$ & \makecell[l]{$(0.0220,0.9990,$\\ \hspace{5ex} $-0.0392,0.0066)$} & \makecell[l]{$(0.0184, 0.9990,$\\ \hspace{5ex} $-0.0394,0.0068)$} & \makecell[l]{$(0.0511,0.9979,$\\ \hspace{5ex} $-0.0386,0.0068)$} & \makecell[l]{$(0.0293,0.9988,$\\ \hspace{5ex} $-0.0390,0.0063)$} \\
\end{tabular}%
}
\caption{Statistics and benchmark information for significant MSSM points marked in Figure~\ref{fig:mssmBinowinoPlot}. The significance and associated signal strength $(p,\hat{\mu})$ are provided for all four analyses for each point. Note that ATLAS monojet sensitivity can only be computed based on the most sensitive individual signal region. All BSM particles not listed here are decoupled to at least 2~TeV.}\label{tab:MSSMsigPointsInfo}
\end{table*}
\renewcommand{\arraystretch}{1.0}

For our analysis toolchain we generate separate samples for the processes
\begin{align}
\nonumber   pp &\to \tilde{\chi}_2^0 \tilde{\chi}_1^\pm,\ \tilde{\chi}_2^0 \rightarrow \ell^+ \ell^- \tilde{\chi}_1^0 \\
\text{and}\ \ \  pp &\to \tilde{\chi}_1^+ \tilde{\chi}_1^-,
\end{align}
as well as a ``monojet sample'' with $pp \rightarrow \tilde{\chi}_2^0 \tilde{\chi}_1^\pm, \tilde{\chi}_1^+ \tilde{\chi}_1^-$ with no forced decay of the neutralino. In each case we allow matrix elements to include up to two additional partons and combine them according to the MLM merging prescription. We generate $3.2\times 10^6$ events per sample per parameter point, of which we keep 110 for all analyses for this model. In similar fashion to the validation of the soft-lepton recasts, the generation is handled by {\sc BSMArt} creating gridpacks from \madgraph, which run on eight cores in parallel, passing their Les Houches Event samples to \hackanalysis for internal showering using \pythia, and finally for analysis. We use the package \textsc{Spey} \cite{Araz:2023bwx} to perform statistical analysis on our samples, including the calculation of expected and observed limits at 95\% confidence level~\cite{Read:2002hq} and of discovery $p$-values.

\subsection{Analysis \& comparison to simplified bino-wino scenario}
\label{s3.2}

The results of our analysis of the bino-wino MSSM are displayed in Figure~\ref{fig:mssmBinowinoPlot} and the accompanying Table~\ref{tab:MSSMsigPointsInfo}. Figure~\ref{fig:mssmBinowinoPlot} shows the expected and observed limits for all searches sensitive to this region of parameter space, as well as a light blue region in which the dark matter relic abundance lies within 20\% of the observed value. It is first illustrative to compare our soft-lepton exclusions to those found for the simplified bino-wino model in the right panels of Figures~\ref{fig:2lValid} and \ref{fig:3lValid}. As expected they are very similar, with the only notable difference being the increased reach of the $2\ell$ search at small $\Delta m$, since we do not include any statistical uncertainty on the signal in the statistics. Meanwhile, we are able to show expected and observed limits from the CMS monojet analysis, CMS-EXO-20-004, but not from its ATLAS equivalent: the latter is much less sensitive and imposes no constraint on the parameter space visible here.

As advertised, both the $2\ell$ and monojet channels are in mild but significant excess. Four statistically significant benchmark points are marked in Figure~\ref{fig:mssmBinowinoPlot} and then detailed in Table~\ref{tab:MSSMsigPointsInfo}. These points are chosen based on their statistics for either the ATLAS $2\ell$ analysis \emph{or} the CMS monojet (\ie, there is no global fit), but we show the discovery significance and associated signal strength with respect to all four analyses for each point. In particular, there is a point with significant discovery $p$-value ($p < 0.05$) and another point with best-fit signal strength $\hat{\mu}$ close to unity for each of the $2\ell$ and (CMS) monojet analyses. Here and throughout the rest of this work, the non-excluded point with $\hat{\mu}$ closest to unity is called ``best-fit point'' for that analysis; the point with the smallest discovery $p$-value is called ``most significant''. For the MSSM, the point at $(m_{\tilde{\chi}^0_2},\Delta m) = (273,16.3)$~GeV, marked as ``ATLAS $2\ell$ best-fit point," has a discovery $p$-value of $p = 0.047$ and an associated signal strength of $\hat{\mu} = 1.12$. It also shows some noteworthy significance, $p < 0.1$, with respect to the CMS monojet analysis, suggesting some overlap between excess regions. The ``ATLAS $2\ell$ most significant point" at $(m_{\tilde{\chi}^0_2},\Delta m) = (284,20.0)$~GeV, meanwhile, has $p = 0.041$ and $\hat{\mu} = 1.26$. The analogous points for the CMS monojet analysis are at $(m_{\tilde{\chi}^0_2},\Delta m) = (287,7.30)$~GeV, with $(p,\hat{\mu}) = (0.049,1.15)$, and $(258,11.8)$~GeV, with $(p,\hat{\mu}) = (0.044,1.40)$. The best-fit point, with smaller $\Delta m$, is essentially invisible to the soft-lepton analyses, but the other point shows some improvement with larger $\Delta m$. We finally comment that the $2\ell$ significant points, particularly the most significant point, lie within striking distance of the band preferred for the dark matter relic abundance, suggesting that this model interpretation, while imperfect, may provide a blueprint for constructing a new model dedicated to fitting these excesses.

%% file: Sections/4_NMSSM.tex
\section{Alternative interpretation: NMSSM with singlino-higgsino LSP}
\label{s4}

The Next-to-Minimal Supersymmetric Standard Model~\cite{Ellwanger:2009dp} (NMSSM) extends the MSSM field content to include a superfield $\mathcal{S}$ that transforms as a singlet under the SM gauge group. A signature feature of this model is the appearance of a superpotential coupling between $\mathcal{S}$ and the Higgs superfields that can take the place of the MSSM $\mu$ term after enforcing a $\mathbb{Z}_3$ symmetry forbidding all gauge-invariant dimensionful superpotential contributions. To be specific, the MSSM superpotential is modified according to 
\begin{align}\label{eq:NMSSMsuperPot}
    \mathcal{W}_{\text{MSSM}} \supset \mu \mathcal{H}_{\text{u}} \mathcal{H}_{\text{d}} \to \mathcal{W}_{\text{NMSSM}} \supset \lambda \mathcal{S} \mathcal{H}_{\text{u}} \mathcal{H}_{\text{d}} + \frac{1}{3}\kappa \mathcal{S}^3,
\end{align}
and the corresponding soft scalar terms go as
\begin{multline}
  V^\mathrm{soft}_{\text{MSSM}} \supset b \mu H_{\text{u}} H_{\text{d}} \to \\ V^\mathrm{soft}_{\text{NMSSM}} \supset A_{\lambda} \lambda\, S H_{\text{u}} H_{\text{d}} + \frac{1}{3} A_{\kappa}\kappa S^3.
\end{multline}
The $\mathcal{S}\mathcal{H}_{\text{u}}\mathcal{H}_{\text{d}}$ term in \eqref{eq:NMSSMsuperPot} generates an effective $\mu$ term when the singlet scalar $S$ obtains a vacuum expectation value (VEV). The NMSSM can thus solve the so-called ``$\mu$ problem'' of the MSSM if $\mu_{\text{eff}} = \lambda \langle S \rangle$ is at the weak scale \cite{Kim:1983dt}. In the CP-conserving limit, the addition of the singlet scalar leads to a physical scalar sector composed of two CP-odd and three CP-even neutral scalars in addition to the familiar charged scalars of the MSSM. Generically, the SM-like Higgs need not be the lightest CP-even scalar and the phenomenology of the NMSSM's scalar sector alone can hence be quite rich. The electroweakino sector is likewise augmented to include an additional neutralino, which can be light.

There is a particularly interesting NMSSM scenario in which the LSP is composed mostly of the singlet fermion (singlino) and $\mu_{\text{eff}}$ is at the weak scale such that the next-lightest electroweakinos are predominantly higgsinos (see also~\cite{Domingo:2022pde,Cao:2022htd,Cao:2023juc} for a discussion of the soft-lepton searches in this context). In this \emph{singlino-higgsino} scenario there are four light electroweakinos $\{\tilde{\chi}^0_1,\tilde{\chi}^0_2,\tilde{\chi}^{\pm}_1,\tilde{\chi}^0_3\}$, nearly aligned with the singlino and triplet of higgsinos. This spectrum opens new channels relevant to the LHC excesses that we are studying, including for instance the process $pp \to \tilde{\chi}^0_2 \tilde{\chi}^0_3$ with both higgsinos decaying to $\tilde{\chi}^0_1$ via $Z$ bosons (which may be off shell if the singlino-higgsino spectrum is sufficiently compressed). On the other hand, production of the LSP itself has quite low cross sections insofar as $\tilde{\chi}^0_1$ is dominated by the singlino. Meanwhile, notably, this scenario also has parameter space predicting the correct dark matter relic abundance through freeze out. The dominant annihilation channels are of singlinos to weak bosons, $\tilde{\chi}^0_1 \tilde{\chi}^0_1 \to W^+W^-/ZZ$, and singlino-higgsino coannihilations to quarks, \eg\ $\tilde{\chi}^0_1 \tilde{\chi}^-_1 \to \bar{u}d$. The contributions of higgsino coannihilations like $\tilde{\chi}^0_{2,3} \tilde{\chi}^-_1 \to \bar{u}d$ are subleading but non-negligible. 

\subsection{Parameter space scan and event generation}
\label{s4.1}

As for our study of the MSSM, we use \textsc{BSMArt} to perform a random scan of the parameter space possibly relevant to the soft-lepton and monojet searches. In the interest of exploring both the collider excesses and the DM relic abundance, we require that the LSP be at least 90\% singlino, which implies the condition $|N_{15}|^2 \geq 0.90$ on the $\tilde{\chi}^0_1$-singlino element of the neutralino mixing matrix. We look for $\{\tilde{\chi}^0_2,\tilde{\chi}^{\pm}_1,\tilde{\chi}^0_3\}$ with similarly high higgsino content. Since these electroweakinos are higgsino-like, we target a similar region of the $(m_{\tilde{\chi}^0_2}, \Delta m \equiv \Delta m(\tilde{\chi}^0_2,\tilde{\chi}^0_1))$ plane as seen in Section \ref{s2} for the ATLAS soft-lepton analyses; namely,
\begin{align}
    m_{\tilde{\chi}^0_2} \in [150,220]~\text{GeV},\ \ \ \Delta m(\tilde{\chi}^0_2,\tilde{\chi}^0_1) \in (0,25]~\text{GeV},
\end{align}
though not quite all of this $\Delta m$ slice is probed in practice. As before, the mass spectrum, mixing matrices, and decays are calculated by \textsc{SPheno}, with fermion (Higgs) masses at one-loop (two-loop) accuracy and decays at tree level. The dark matter relic abundance is computed by \textsc{MicrOMEGAs}, including LSP coannihilations and NLSP (co)annihilations, and the usual range of experimental constraints are imposed by \textsc{HiggsTools} and \textsc{SModelS} for points appearing to inhabit the soft-lepton/monojet excess region. The lightest CP-even scalar is required to have mass $m_h \in [122,128]$~GeV.

\begin{figure}
    \centering
    \includegraphics[scale=0.7]{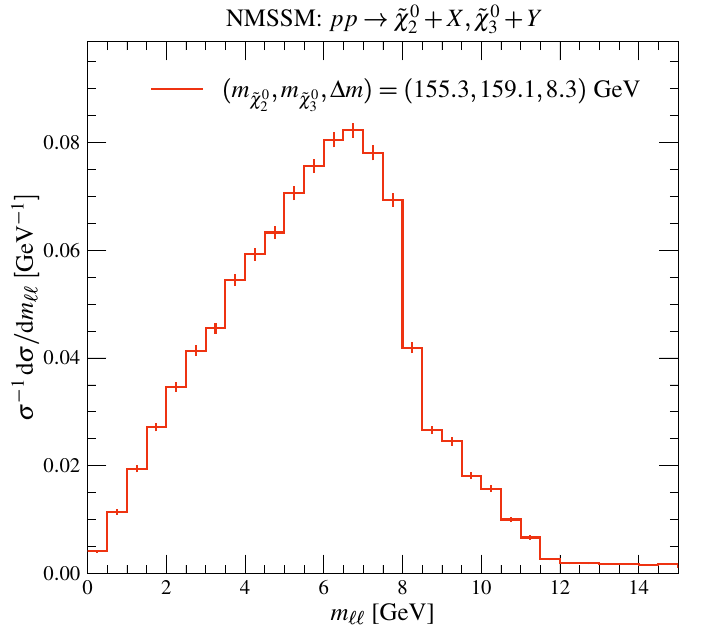}
    \caption{$m_{\ell\ell}$ distribution for the NMSSM with a singlino-like LSP as found from our simulations for the processes $p p \rightarrow \tilde{\chi}^0_2 +X$ and $pp \to \tilde{\chi}^0_3 + Y$, where $X \in \{\tilde{\chi}_1^\pm,\tilde{\chi}_3^0\}$ and $Y \in \{\tilde{\chi}_1^\pm,\tilde{\chi}_2^0\}$, with matrix elements including up to two additional partons. Here $\Delta m = m_{\tilde{\chi}^0_2} - m_{\tilde{\chi}^0_1}$, as for the MSSM.}\label{fig:nmssmc1nimll}
\end{figure}

\begin{figure}
    \centering
    \includegraphics[scale=0.7]{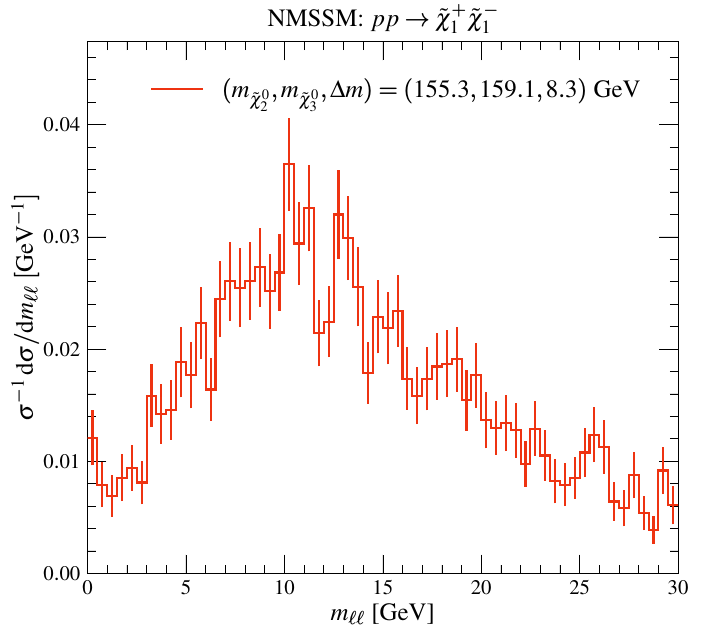}
    \caption{$m_{\ell\ell}$ distribution for the NMSSM with a singlino-like LSP as found from our simulations for the chargino pair-production process $p p \rightarrow \tilde{\chi}_1^+ \tilde{\chi}_1^-$, with matrix elements including up to two additional partons.}\label{fig:nmssmc1c1mll}
\end{figure}

For this scan we again run Fortran code created with \textsc{SARAH} linked to the \textsc{SPheno} library, with a low-scale input card. We solve the tadpole equation for the singlet scalar in order to use $\mu_{\text{eff}}$ instead of the singlet VEV $\langle S \rangle$ as an input. The full set of variable inputs includes the Higgs sector parameters
\begin{align}
\nonumber    \mu_{\text{eff}} &\in [100,250]~\text{GeV},\\
\nonumber    \kappa &\in [1,20] \times 10^{-3},\\
\nonumber    \delta &\in [0.01,0.20]\ \ \ \text{such that}\ \ \  \lambda = (2 + \delta)\kappa,\\
\tan \beta &\in [8,20]
\end{align}
and the soft (s)fermion SUSY-breaking parameters
\begin{align}
\nonumber    M_{\tilde{t}}^2 &\in [2.5, 10] \times 10^7~\text{GeV}^2,\\
\nonumber    A_t &\in [-5,5] \times 10^3~\text{GeV},\\
\nonumber    A_{\lambda} &\in [-3,3] \times 10^2~\text{GeV},\\
A_{\kappa} &\in [-3,3] \times 10^2~\text{GeV};
\end{align}
while the other relevant parameters are fixed as follows:
\begin{align}
  \!\!  M_1 = M_2 &= 2~\text{TeV},\\
\nonumber  \!\!  M_{\tilde{q}\neq \tilde{t}}^2 &= (8.25 \times 10^8, 8.25 \times 10^8)~\text{GeV}^2,\\ 
\nonumber    M_{\tilde{u}}^2 = M_{\tilde{d}}^2 &= (8.25 \times 10^8, 8.25 \times 10^8, 1.25 \times 10^7)~\text{GeV}^2,\\
    \!\! M_{\tilde{\ell}}^2 = M_{\tilde{e}}^2 &= (4.25 \times 10^8, 4.25 \times 10^8, 4.25 \times 10^6)~\text{GeV}^2.\nonumber
\end{align}

\begin{figure*}
\centering
\includegraphics[width=.7\textwidth]{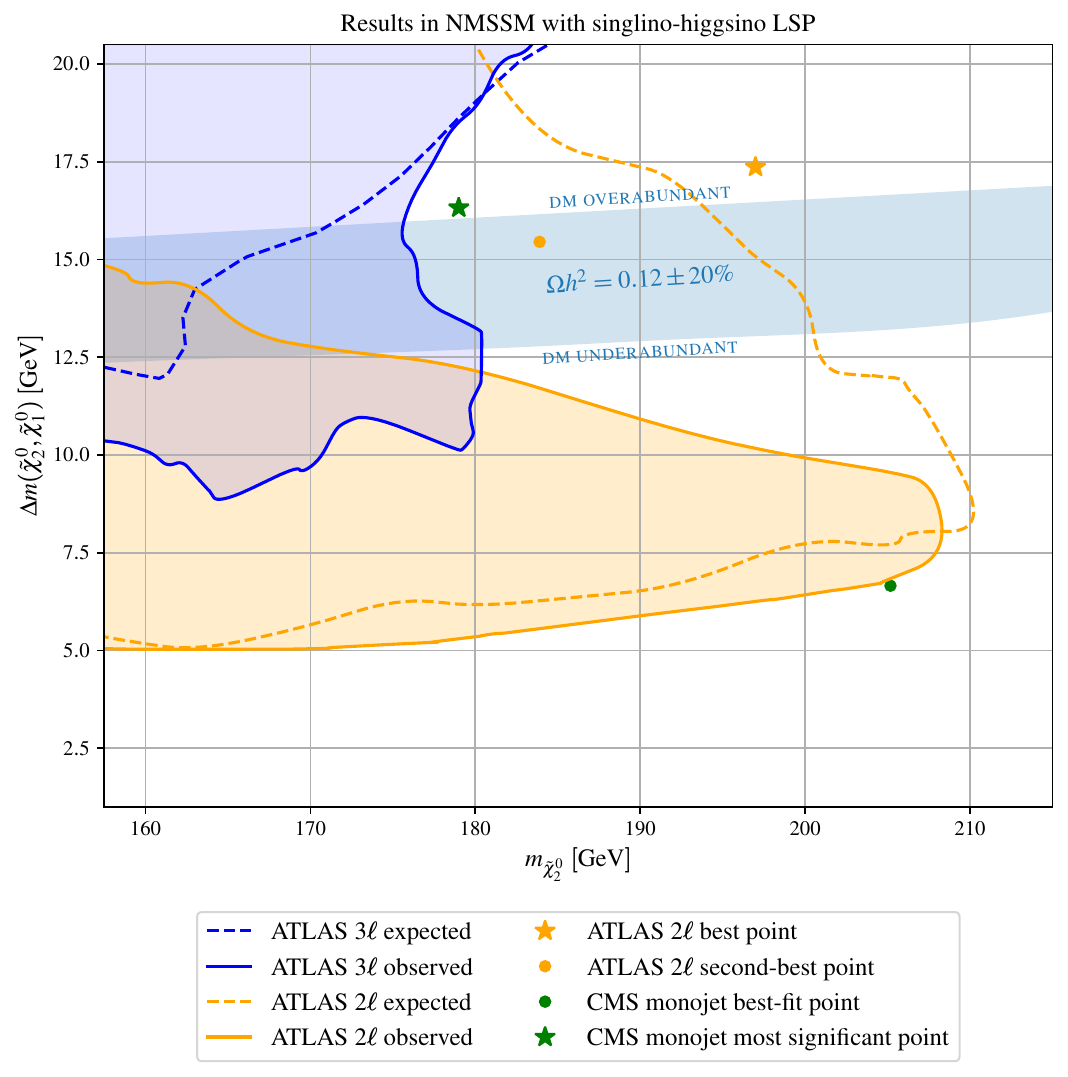}
\caption{View of NMSSM parameter space relevant to the LHC excesses. Orange contours show limits from ATLAS-SUSY-2018-16 ($2\ell$), while blue contours refer to ATLAS-SUS-2019-09 ($3\ell$). Dashed contours show expected limits and solid contours denote the edge of the observed excluded regions, shaded correspondingly. The light blue band indicates approximately the region of the parameter space with relic abundance $\Omega h^2$ within 20\% of the \emph{Planck} measurement. Two interesting non-excluded points are highlighted for each of the $2\ell$ and monojet analyses.}\label{fig:NMSSMexclusionPlot}
\end{figure*}The requirement that $\lambda \gtrsim 2\kappa$ arises from the need for a singlino-like LSP and $\mathcal{O}(10)$~GeV compression between the singlino-dominated state and the higgsinos. The relic abundance turns out to be sensitive to the deviation $\delta$ from the exact relation $\lambda = 2\kappa$, with $\delta \approx 0.07$ producing the majority of points with $\Omega h^2 \approx 0.12$. Our collection includes points with larger $\delta$, which have overabundant DM with relic densities as high as $\Omega h^2 = 0.2$, in order to better cover the range of $\Delta m$ values relevant to the LHC discussion.

\renewcommand{\arraystretch}{1.1}\setlength{\tabcolsep}{12pt}
\begin{table*}
\centering
\resizebox{0.98\textwidth}{!}{%
    \begin{tabular}{l l@{\hspace{3ex}}l@{\hspace{3ex}}l@{\hspace{3ex}}l}
Point & {\color{mpl-orange}$\bigstar$} ATLAS $2\ell$ best & {\color{mpl-orange}\scalebox{1.2}{$\bullet$}} ATLAS $2\ell$ second best & {\color{mpl-green}\scalebox{1.2}{$\bullet$}} CMS monojet best fit & {\color{mpl-green}$\bigstar$} CMS monojet most significant \\
$(m_{\tilde{\chi}^0_2},\Delta m)$ [GeV]\ \ \ \ \ & (197,\,17.4) & (184,\,15.5) & (205,\,6.66) & (179,\,16.3) \\
\midrule
$(p,\hat{\mu})$ [ATLAS $2\ell$] & (0.041,\,0.97) & (0.044,\,0.80) & $(0.435,\,<0.1)$ & (0.071,\,0.64) \\
$(p,\hat{\mu})$ [ATLAS $3\ell$] & (0.446,\,0.12) & $(>0.5,\,0.12)$ & $(>0.5,\,0.71)$ & $(>0.5,\,0.13)$ \\
$(p,\hat{\mu})$ [CMS monojet] & (0.132,\,3.00) & (0.129,\,2.65) & (0.712,\,1.91) & (0.051,\,3.79) \\
$(p,\hat{\mu})$ [ATLAS monojet] & (0.277,\,2.44) & (0.277,\,2.02) & (0.127,\,2.96) & (0.277,\,2.08) \\
\midrule
$\mu_{\text{eff}}$ [GeV] & 189.3 & 177.0 & 199.1 & 172.6 \\
$\kappa$ & 0.0157 & 0.0108 & 0.0025 & 0.0146 \\
$\lambda$ & 0.0330 & 0.0226 & 0.0050 & 0.0309 \\
$\tan \beta$ & 19.71 & 25.94 & 10.70 & 12.82 \\
$M_{\tilde{t}}^2$ [GeV$^2$] & $8.06 \times 10^7$ & $7.20 \times 10^7$ & $3.42 \times 10^7$ & $9.12 \times 10^7$\\
$A_t$ [GeV] & $2.61 \times 10^3$ & $-1.28 \times 10^3$ & $2.07 \times 10^3$ & $-2.64 \times 10^3$\\
$A_{\lambda}$ [GeV] & $-34.60$ & $-92.77$ & 189.4 & 192.0 \\
$A_{\kappa}$ [GeV] & $-43.01$ & $-8.771$ & $-161.3$ & $-55.91$ \\
\midrule
$m_h$ [GeV] & 124.0 & 123.4 & 123.0 & 122.6 \\
$m_{\tilde{\chi}^0_1}$ [GeV] & 179.6 & 168.5 & 198.5 & 162.7 \\
$m_{\tilde{\chi}^0_2}$ [GeV] & 197.0 & 183.9 & 205.2 & 179.0 \\
$m_{\tilde{\chi}^{\pm}_1}$ [GeV] & 198.1 & 185.5 & 207.1 & 180.3 \\
$m_{\tilde{\chi}^0_3}$ [GeV] & 199.9 & 187.3 & 209.0 & 182.1 \\
$(N_{11},N_{12},N_{13},N_{14},N_{15})$ & \makecell[l]{$(0.0042,-0.0070,$\\ \hspace{3ex} $0.1547,-0.1683,$\\ \hspace{13ex} $0.9735)$} & \makecell[l]{$(0.0032,-0.0053,$\\ \hspace{3ex} $0.1201,-0.1299,$\\ \hspace{13ex} $0.9841)$} & \makecell[l]{$(0.0016,-0.0026,$\\ \hspace{3ex} $0.0580,-0.0597,$\\ \hspace{13ex} $0.9965)$} & \makecell[l]{$(0.0041,-0.0069,$\\ \hspace{3ex} $0.1479,-0.1622,$\\ \hspace{13ex} $0.9756)$} \\
$(N_{21},N_{22},N_{23},N_{24},N_{25})$ & \makecell[l]{$(-0.0173,0.0289,$\\ \hspace{3ex} $-0.6932,0.6827,$\\ \hspace{13ex} $0.2284)$} & \makecell[l]{$(-0.0172,0.0287,$\\ \hspace{3ex} $-0.7004,0.6907,$\\ \hspace{13ex} $0.1767)$} & \makecell[l]{$(-0.0184,0.0312,$\\ \hspace{3ex} $-0.7077,0.7006,$\\ \hspace{13ex} $0.0832)$} & \makecell[l]{$(-0.0176,0.0294,$\\ \hspace{3ex} $-0.6951,0.6838,$\\ \hspace{13ex} $0.0219)$} \\
$\Im\,(N_{31},N_{32},N_{33},N_{34},N_{35})$ & \makecell[l]{$(-0.0134,0.0226,$\\ \hspace{3ex} $0.7039,0.7097,$\\ \hspace{13ex} $0.0110)$} & \makecell[l]{$(-0.0137,0.0231,$\\ \hspace{3ex} $0.7036,0.7101,$\\ \hspace{13ex} $0.0080)$} & \makecell[l]{$(-0.0126,0.0216,$\\ \hspace{3ex} $0.7041,0.7097,$\\ \hspace{13ex} $0.0016)$} & \makecell[l]{$(-0.0131,0.0220,$\\ \hspace{3ex} $0.7035,0.7100,$\\ \hspace{13ex} $0.0116)$} \\
\end{tabular}%
}
\caption{Statistics and benchmark information for significant NMSSM points marked in Figure~\ref{fig:NMSSMexclusionPlot}. The significance and associated signal strength $(p,\hat{\mu})$ are provided for all four analyses for each point. Neutralino masses are forced positive such that the $\tilde{\chi}^0_3$ mixing matrix elements $N_{3i}$ (with $i=1,2,\ldots,5$) are imaginary.}\label{tab:NMSSMsigPointsInfo}
\end{table*}
\renewcommand{\arraystretch}{1.0}

For our analysis toolchain, in a similar way as for the MSSM points, we generate three separate samples involving off-shell $Z$ decays to leptons; purely off-shell $W$ decays to leptons; and a separate monojet sample with flat ($|\mathcal{M}|^2 \to 1$) decays handled by \pythia. As mentioned above, the higgsino-like $\tilde{\chi}_2^0$ and $\tilde{\chi}_3^0$ are both light and can decay via off-shell $Z$ bosons, which leads to a complicated set of processes. This is interesting in itself: in the basis where the neutralino mixing matrix is real, it turns out that our $\tilde{\chi}^0_2$ has positive mass (as does $\tilde{\chi}^0_1$) while $\tilde{\chi}^0_3$ has the opposite. In keeping with the discussion in Section \ref{s2.1}, this spectrum suggests that $\tilde{\chi}^0_2$ decays with an $m_{\ell\ell}$ distribution similar to the ATLAS wino/bino(+) scenario, while $\tilde{\chi}^0_3$ decays like a higgsino. This expectation is borne out in Figure~\ref{fig:nmssmc1nimll} for a typical point produced by our scan: by comparison with Figure~\ref{fig:mllcomparison}, one can make out in Figure~\ref{fig:nmssmc1c1mll} the linear combination of $\tilde{\chi}^0_2$ and $\tilde{\chi}^0_3$ $m_{\ell\ell}$ distributions. Due to the complexity of the multicomponent signal, we use the leading-order cross sections reported by \madgraph and \pythia after MLM merging (with merging scale set again to $40$ GeV). We perform simulations for 194 parameter points for the soft-lepton analyses. The generation of the monojet sample(s) involves the production of all possible pairs of light electroweakinos, and relies on matrix elements featuring up to two additional partons. It is worth noting that the $p p \rightarrow \tilde{\chi}_1^+ \tilde{\chi}_1^-$ samples are relatively simple and the subsequent $m_{\ell\ell}$ distribution, shown for our example point in Figure~\ref{fig:nmssmc1c1mll}, is similar to the chargino-pair distributions for the MSSM shown in Figure~\ref{fig:higgsinoc1c1mll}. For the monojet analyses we simulate 238 parameter points. As before, we generate $3.2\times 10^6$ events per sample per parameter point. The generation is handled by {\sc BSMArt} creating gridpacks from \madgraph, which run on eight cores in parallel, passing their Les Houches Event samples to \hackanalysis for showering, analysis, and statistics. Our simulation relies on a UFO model produced by \textsc{SARAH}.

\subsection{Analysis: Degraded monojet fit}
\label{4.2}

The results of our NMSSM analysis are displayed in Figure~\ref{fig:NMSSMexclusionPlot} and the accompanying Table~\ref{tab:NMSSMsigPointsInfo}. Figure~\ref{fig:NMSSMexclusionPlot} shows the expected and observed limits from the soft-lepton analyses, which are the only searches sensitive to this part of the $(m_{\tilde{\chi}^0_2},\Delta m)$ plane. In particular, both monojet analyses are markedly less sensitive to this model than to the decoupled-higgsino MSSM; neither search excludes any of the points in our scan. This reduced sensitivity degrades the overlap between the monojet excesses (which do still appear in the form of weaker observed limits on the monojet signal strength) and the $2\ell$ excess, and causes strong tension between the monojet discovery $p$-values and best-fit signal strengths. Figure~\ref{fig:NMSSMexclusionPlot} also shows that, unlike in the MSSM, the observed limit from ATLAS-SUSY-2019-09 is mostly stronger than the expected limit. 

Despite the lack of a limit from either monojet search, we report four statistically significant benchmark points in Table~\ref{tab:NMSSMsigPointsInfo}. As for the MSSM, we report the discovery significance and associated signal strength for all four analyses for each point. The non-excluded point with the greatest significance with respect to the $2\ell$ analysis is at $(m_{\tilde{\chi}^0_2},\Delta m) = (197,17.4)$~GeV; it carries a discovery $p$-value of $p = 0.041$ associated with a signal strength of $\hat{\mu} = 0.97$. Here already we can see the tension with the CMS monojet analysis, which requires more than thrice our yield for meaningful discovery potential. Another interesting point with high significance lies at $(m_{\tilde{\chi}^0_2},\Delta m) = (183.9,15.5)$~GeV, with $p = 0.044$ and $\hat{\mu} = 0.80$. The fit at this point is imperfect, but it lies within the band predicting a reasonable DM relic abundance. Meanwhile, for the CMS monojet analysis, the non-excluded point\footnote{We are now discussing the sensitivity of the monojet analysis, but the limiting search is invariably one of the soft-lepton analyses, so by ``non-excluded'' we mean allowed by one of those.} with the best fit\footnote{For the NMSSM monojet results, we require $\hat{\mu}<4$ in order to limit ourselves to points with somewhat reasonable fits.} is at $(m_{\tilde{\chi}^0_2},\Delta m) = (205,6.66)$~GeV and has $\hat{\mu} = 1.914$ with $p = 0.712$. This point is a poor fit for both soft-lepton excesses. The (non-excluded) monojet point with greatest significance, on the other hand, is at $(m_{\tilde{\chi}^0_2},\Delta m) = (179,16.3)$~GeV; its discovery $p$-value is $p = 0.051$, but the associated signal strength is $\hat{\mu} = 3.791$, suggesting a severe deficit of monojet events at this point. Altogether, we find a strong contrast between the global picture in the bino-wino MSSM (which shows some overlap between statistically significant monojet and $2\ell$ excesses) and the NMSSM, which exhibits a better fit between the $2\ell$ excess and the DM relic abundance at the expense of the monojet fit.

%% file: Sections/5_Non_SUSY.tex
\section{Non-supersymmetric interpretations}
\label{s5}

Having shown that two supersymmetric scenarios give fits of varying quality to the soft-lepton and monojet excesses, we now consider whether some non-supersymmetric model might be compatible with either or both. Two well motivated frameworks capable of producing events with the correct final states include scalar dark matter coupling to the Standard Model with a vector-like lepton~\cite{Bai:2014osa} and a Higgs triplet model originally proposed to generate neutrino masses~\cite{Konetschny:1977bn, Cheng:1980qt, Lazarides:1980nt, PhysRevD.22.2227, Mohapatra:1980yp, Cai:2017mow}. Here we briefly review each scenario and describe our scans over their relevant parameter spaces.

\subsection{Scalar dark matter with vector-like leptons}

\begin{figure}
\centering
\includegraphics[scale=1]{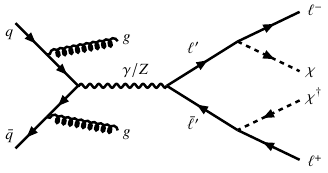}
\caption{\label{fig:VLLlep1}Soft-lepton signal from pair production of a vector-like charged lepton $\ell'$ at the LHC, followed by the direct decay of each $\ell'$ to a SM lepton $\ell$ and the DM particle $\chi$.}
\end{figure}

\begin{figure}\label{vectorLikeFigs}
\centering
\includegraphics[scale=1]{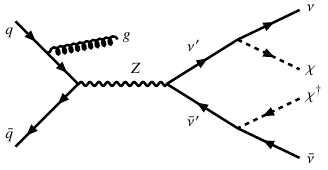}
\caption{\label{fig:VLLmonojet}Monojet signal from pair production of the vector-like neutrino $\nu'$ at the LHC, followed by the direct decay of each $\nu'$ to a SM neutrino $\nu$ and the DM particle $\chi$.}
\end{figure}

One class of simplified dark matter models features scalar or fermionic DM coupling directly to a SM lepton and a mediating field with identical SM gauge quantum numbers \cite{Bai:2014osa}. We focus on scenarios with scalar dark matter $\chi$ interacting with vector-like mediating lepton(s) (VLLs). In principle, the dark matter could couple to right- or left-handed SM leptons along with some suitable new VLL field(s). In the first case, the VLL would have SM quantum numbers $(\boldsymbol{1},\boldsymbol{1},-1)$; in the second case it must lie in the representation $(\boldsymbol{1},\boldsymbol{2},-\tfrac{1}{2})$. In the singlet VLL case, events with an opposite-charge lepton pair, missing energy, and jets can result from VLL pair production and decay, as depicted in Figure \ref{fig:VLLlep1}. The invariant mass of the $\ell^+\ell^-$ pair can be low if the mass splitting $\Delta m(\ell',\ell)$ is small, of $\mathcal{O}(1\text{--}10)$~GeV. On the other hand, a monojet signature is unlikely to arise in this model unless the SM leptons are too soft to be observed. The doublet VLL model, on the other hand, can produce \emph{both} soft-lepton events in the same topology as the singlet VLL model \emph{and} monojet events via pair production of the electrically neutral $\nu'$, as shown in Figure \ref{fig:VLLmonojet}. Therefore we expect that the doublet VLL model may be the better choice for the excesses in question, and we focus on this scenario in our analysis. For a single flavor of doublet VLL, the most general DM-VLL interaction Lagrangian can be written as
\begin{align}\label{eq:VLLLag}
    \mathcal{L} \supset \lambda_I\, \chi\bar{\Psi}L_I + \text{H.c.},
\end{align}
where $\Psi$ denotes a weak doublet of vector-like leptons and $L_I$ with $I \in \{1,2,3\}$ is a SM lepton weak doublet, with the parameter $\lambda$ representing a vector of VLL couplings in the flavor space. Regarding global symmetries, we suppose that (despite the name) the VLL fields do not carry lepton number. Lepton number is instead conserved because the complex scalar $\chi$ carries $L = -1$. We therefore neglect mixing between the VLL and any SM lepton. Meanwhile, if the dark matter and VLL(s) are odd under a $\mathbb{Z}_2$ symmetry under which all SM fields have even parity, then $\chi$ is a viable dark matter candidate as long as $\chi$ is lighter than the VLL(s), as in all $t$-channel DM simplified models.

 To fix notation, let the VLL doublet have components denoted by $\Psi = (\nu',\ell')^{\transpose}$ in analogy with the SM doublets $L_I = (\nu_{\text{L}I},\, \ell_{\text{L}I})^{\transpose}$. The operator \eqref{eq:VLLLag} then produces the couplings
\begin{align}\label{eq:VLLdoublet}
    \mathcal{L} \supset \lambda_I\,\chi(\bar{\nu}{}' \nu_{\text{L}I} + \bar{\ell}'\ell_{\text{L}I}) + \text{H.c.}
\end{align}
Meanwhile, the $\Psi$ kinetic terms include $\bar{\nu}{}'\nu'/\bar{\ell}{}'\ell'(\gamma/Z)$ and $\bar{\nu}{}'\ell'W^+$ interactions. The latter coupling seems interesting because it provides a distinct decay channel for one component of $\Psi$ that can produce a soft lepton through its decay via an off-shell $W$ boson, but in practice this $W$-mediated three-body decay is always dominated by the Yukawa-like two-body decays from \eqref{eq:VLLdoublet}.

It is of course interesting to consider the phenomenology of the dark matter candidate $\chi$. Its annihilation is dominated by $t$-channel diagrams to pairs of SM leptons and neutrinos, mediated by their heavy partners, but in the parameter space regions that we consider, coannihilations of $\chi$ with the vector-like particles and annihilations of the VLLs themselves are quite efficient. Many processes in the latter category are moreover proportional to gauge couplings and thus cannot be controlled by our choices of Yukawa-like couplings $\lambda_I$. Therefore, as we show below, not only does the \emph{Planck} relic abundance favor quite small values of $\lambda_I$ (of $\mathcal{O}(10^{-3})$, which seems to escape direct-detection constraints \cite{Bai:2014osa,Iguro:2022tmr}), but in fact the \emph{Planck} value seemingly cannot be attained for arbitrarily small mass splittings between the VLLs and the dark matter.

\begin{figure}
    \centering
    \includegraphics[scale=0.7]{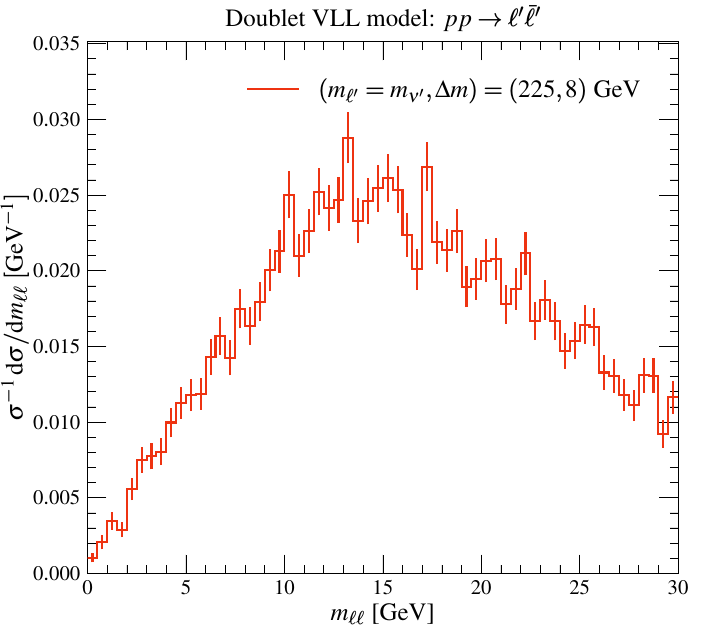}
    \caption{$m_{\ell\ell}$ distribution resulting from the production and decay of a pair of doublet vector-like leptons, $p p \rightarrow \ell' \bar{\ell}{}' \to \ell^+\ell^-\chi\chi$, after combining matrix elements featuring up to two additional partons. We consider a scenario with a vector-like lepton mass of $225$ GeV and dark matter scalar mass of $217$ GeV.}\label{fig:leptondm2mll}
\end{figure}

We perform an enhanced scan with $9.6 \times 10^6$ events per point, simulating pair production of vector-like leptons relying on matrix elements including up to two additional partons. Event generation is performed by \madgraph using a UFO module produced by \frules \cite{Christensen:2009jx, Alloul:2013bka} based on our own implementation of the doublet VLL model. In order to run the recast of the monojet search at the same time, we simulate $p p \rightarrow \bar{\ell}{}^\prime \nu^\prime$ and $p p \rightarrow \bar{\nu}{}^\prime \nu^\prime$ events in the same samples. We fix the branching ratios of $\ell^\prime$ to be $50\%$ to electrons plus dark matter, $50\%$ to muons plus dark matter; and of $\nu^\prime$ entirely to neutrinos plus dark matter. We first performed simulations for 131 parameter points in this model, which sufficed for the soft-lepton analyses, and added 44 more points in order to improve coverage of the $(m_{\ell'},\Delta m)$ plane for the monojet analyses. In Figure~\ref{fig:leptondm2mll} we show the $m_{\ell\ell}$ distribution of events in our data, which shows a peak at roughly $2 \Delta m$ but is much broader and with a longer tail than for the three-body decay processes that we have considered so far. We finally note that the dark matter relic abundance is computed for these benchmark points at leading order by \textsc{MadDM} \cite{Backovic:2013dpa, Ambrogi:2018jqj, Arina:2021gfn}, which is convenient since it obviates the need to produce a separate input for \textsc{MicrOMEGAs}. Since the VLL branching fractions are fixed to the desired values simply by setting $\lambda_1 = \lambda_2$ and their size is irrelevant for the collider study, these are fixed for event generation but allowed to vary as we search for parameter space predicting the correct DM relic abundance.

\begin{figure*}
\centering
\includegraphics[width=.7\textwidth]{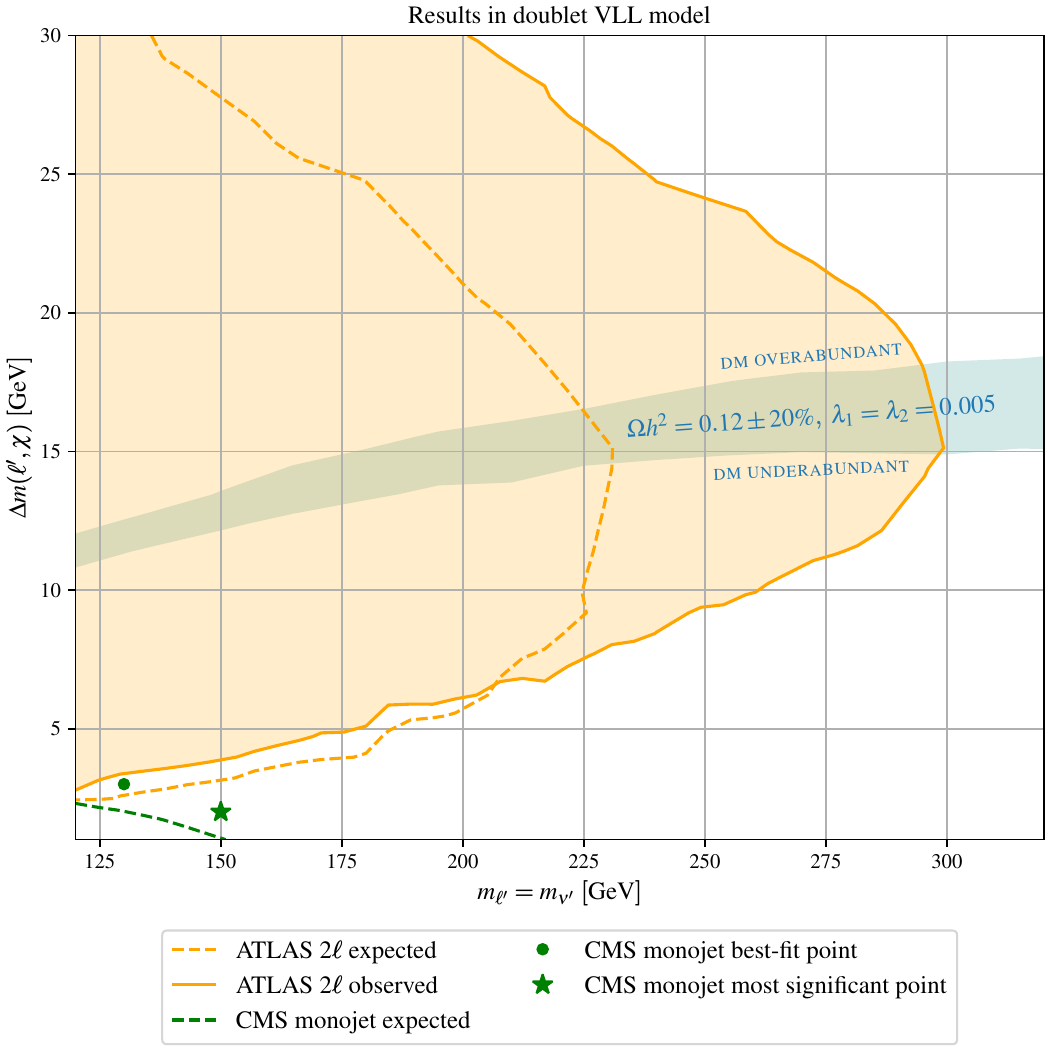}
\caption{View of the region of the VLL doublet model parameter space relevant to the LHC excesses. Orange contours show limits from ATLAS-SUSY-2018-16 ($2\ell$), while green contour refers to to CMS-EXO-20-004 (monojet). Dashed contours indicate expected limits and solid contours denote the edge of the observed excluded regions of the parameter space, shaded correspondingly. Light blue band shows the region approximately featuring a relic abundance $\Omega h^2$ within 20\% of the \emph{Planck} measurement. Two interesting non-excluded points are highlighted for the monojet analysis.}\label{fig:lepDMDoubletexclusionPlot}
\end{figure*}

The results of our scan of the VLL doublet model are displayed in Figure \ref{fig:lepDMDoubletexclusionPlot} and the accompanying Table~\ref{tab:lepDMsigPointsInfo}. Figure~\ref{fig:lepDMDoubletexclusionPlot} is quite different from Figures \ref{fig:mssmBinowinoPlot} and \ref{fig:NMSSMexclusionPlot}, which reflects the unique physics of this non-supersymmetric model. We first see that the ATLAS $2\ell$ search is not at all in excess for this model; accordingly, there are no points with high discovery significance for that analysis. (The $3\ell$ analysis is insensitive to this model by construction: there are no processes producing three leptons.) This behavior is at least partially due to the $m_{\ell\ell}$ distributions characteristic of this model: as noted above, the vector-like leptons tend to produce leptons hard enough to populate bins of the ATLAS-SUSY-2018-16 analysis in which underfluctuations are reported, and which by themselves can powerfully exclude models predicting even a few high-$m_{\ell\ell}$ events.

On the other hand, the monojet excess does survive for this model, so we report two statistically significant points associated with this analysis in Table~\ref{tab:lepDMsigPointsInfo}. There is a point with good significance for the CMS monojet search at $(m_{\ell'},\Delta m(\ell',\chi)) = (150,2)$~GeV, which has $p = 0.0322$ and $\hat{\mu} = 1.097$. There is a nearby point with a slightly better fit at $(m_{\ell'},\Delta m) = (130,3)$~GeV; it has $p = 0.0454$ and $\hat{\mu} = 1.071$. For illustrative purposes, as for previous models, we include in Figure \ref{fig:lepDMDoubletexclusionPlot} a band in which the dark matter relic abundance (computed at leading order by \textsc{MadDM}) lies within 20\% of the observed value by the Planck Collaboration for a benchmark with $\lambda_1 = \lambda_2 = 5 \times 10^{-3}$. As noted above, it is not possible to achieve the correct relic abundance in the vicinity of the monojet points. Coannihilations with $\ell'$ and $\nu'$, most of which are of electroweak strength and independent of the Yukawa-like couplings $\lambda_I$, are too efficient when the mass splittings are of $\mathcal{O}(1)$~GeV, rendering $\chi$ underabundant. We note that the scalar DM annihilation cross section is significantly enhanced at next-to-leading order by three-body processes featuring the emission of a photon (\eg, $\chi\chi \to \ell^+\ell^- \gamma$); the NLO corrections to our results, in this regime where $m_{\ell'}/m_{\chi} \approx (1.0,1.25)$, can be of $\mathcal{O}(10)\%$ \cite{Toma:2013bka,Giacchino:2013bta}.

\renewcommand{\arraystretch}{1.1}\setlength{\tabcolsep}{12pt}
\begin{table*}
\centering
    \begin{tabular}{l l@{\hspace{3ex}}l}
Point & {\color{mpl-green}\scalebox{1.2}{$\bullet$}} CMS monojet best fit & {\color{mpl-green}$\bigstar$} CMS monojet most significant \\
$(m_{\ell'}=m_{\nu'},\Delta m(\ell',\chi))$ [GeV]\ \ \ \ \ & (273,\,16.3) & (284,\,20.0) \\
\midrule
Significance ($p$-value) & 0.047 & 0.041 \\
Signal strength $\hat{\mu}$ & 1.12 & 1.26 \\
\end{tabular}
\caption{Statistics and benchmark information for the significant doublet VLL model points marked in Figure~\ref{fig:lepDMDoubletexclusionPlot}.}\label{tab:lepDMsigPointsInfo}
\end{table*}
\renewcommand{\arraystretch}{1.0}

\subsection{Additional scalars in the type-II seesaw model}

\begin{figure}
\centering
\includegraphics[scale=1]{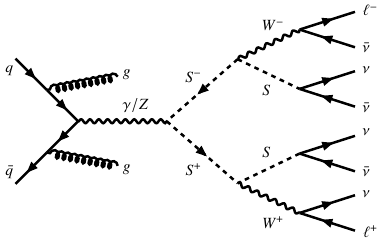}
\caption{\label{fig:type2ewino}Soft-lepton signal from pair production of a singly charged scalar $S^{\pm}$ at LHC, followed by decays to neutral scalars $S$, themselves decaying invisibly.}
\end{figure}

Another longstanding problem, the origin of neutrino masses, is addressed by the type-II seesaw model \cite{Konetschny:1977bn, Cheng:1980qt, Lazarides:1980nt, PhysRevD.22.2227, Mohapatra:1980yp, Cai:2017mow}, which provides a UV completion of the Weinberg operator
\begin{align}
    \mathcal{L}_{\text{Weinberg}} = \epsilon_{ij}\epsilon_{kl}\, \overbar{L^{\text{c}}}_i L_j \Phi_k \Phi_l,
\end{align}
where $\Phi$ and $L$ represent the SM Higgs and lepton doublets~\cite{PhysRevLett.43.1566}. The type-II seesaw model adds to the Standard Model a $Y=1$ weak-triplet scalar $\Delta$ that can be parameterized as
\begin{align}
    \Delta = \frac{1}{\sqrt{2}}\begin{pmatrix}
        \Delta^+ & \sqrt{2} \Delta^{++}\\
        \sqrt{2} \Delta^0 & -\Delta^+
    \end{pmatrix}.
\end{align}
The scalar potential is given by
\begin{multline}
    V(\Phi,\Delta) = -m_{\Phi}^2 \Phi^{\dagger}\Phi + \frac{1}{4}\lambda(\Phi^{\dagger}\Phi)^2 + m_{\Delta}^2 \tr \Delta^{\dagger}\Delta \\ + (\mu\, \Phi^{\transpose} \ii \sigma^2 \Delta^{\dagger} \Phi + \text{H.c.}) + \lambda_1 \Phi^{\dagger}\Phi \tr \Delta^{\dagger}\Delta\\ + \lambda_2 (\tr \Delta^{\dagger}\Delta)^2 + \lambda_3 \tr\, (\Delta^{\dagger}\Delta)^2 + \lambda_4 \Phi^{\dagger} \Delta \Delta^{\dagger}\Phi.
\end{multline}
The physical scalars comprise two CP-even scalars $h, S$, two CP-odd scalars $G^0,A$ ($G^0$ becoming the longitudinal mode of the $Z$ boson), two singly charged scalars $G^{\pm},S^{\pm}$ ($G^{\pm}$ becoming the longitudinal mode of $W^{\pm}$), and a doubly charged scalar $S^{\pm\pm}$. All physical states except the doubly charged scalar are mixtures of the components of the SM-like weak doublet $\Phi$ and of the triplet $\Delta$.
It has been pointed out \cite{PhysRevD.106.075028} that relatively light charged scalars in this model are experimentally allowed when the mixing between $\Delta$ and $\Phi$ is small (so that the charged scalars decay predominantly via $W$ bosons) and when the hierarchy of CP-even scalars is $\{S^{\pm\pm},S^{\pm},S\}$ in order of decreasing mass. At tree level, the mass splittings between scalars satisfy the relation
\begin{align}
    m_{S^{\pm\pm}}^2 - m_{S^{\pm}}^2 = m_{S^{\pm}}^2 - m_S^2 = -\frac{1}{4}\lambda_4v^2,
\end{align}
so that $S^{\pm\pm}$ is the heaviest scalar only if $\lambda_4 < 0$. Meanwhile, if the triplet VEV $v_{\Delta}$ is small enough, one or both of the electrically neutral scalars $\{S,A\}$ decays almost entirely to neutrinos. Thus in scenarios exhibiting small scalar mixing and low $v_{\Delta}$, the type-II seesaw model can generate events with soft leptons and missing energy through scalar pair production. An example process is shown in Figure \ref{fig:type2ewino}. But it is immediately apparent that pair production of the doubly charged scalar may produce additional soft jets or leptons that may be rejected by the search criteria.

\begin{figure}
    \centering
    \includegraphics[scale=0.7]{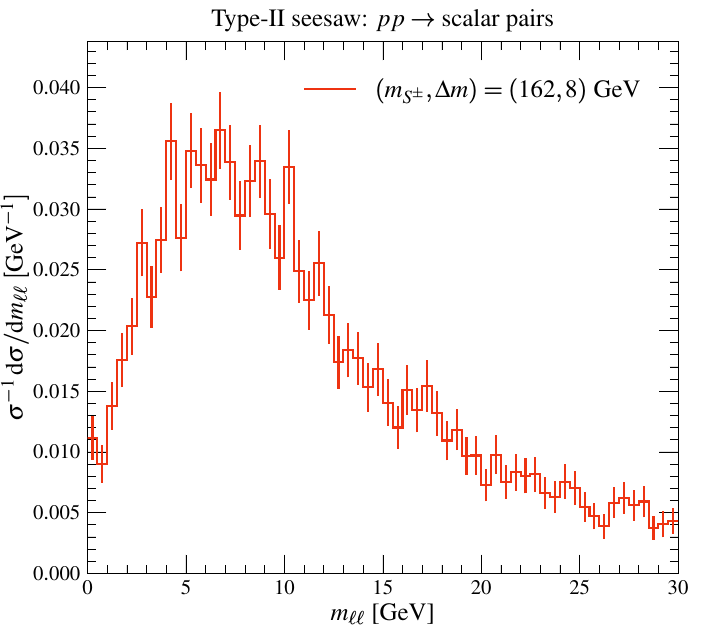}
    \caption{$m_{\ell\ell}$ distribution for the type-II seesaw model, for $m_{S^{\pm}} = 162$ GeV. Processes considered are listed in \eqref{eq:seesawProcesses} and our simulation relies on matrix elements including up to two additional partons. Here $\Delta m = m_{S^{\pm\pm}}- m_{S^{\pm}} = m_{S^{\pm}} - m_{S}$.}\label{fig:typeIImll}
\end{figure}

We perform a scan over $m_{S^{\pm\pm}} \in [100,200]$ GeV and $\Delta m \in [1,30]$ GeV. As usual, we use the \madgraph package; in this case the input UFO is produced using the publicly available \frules implementation of the type-II seesaw model \cite{Fuks:2019clu}. We simulate all at once the processes
\begin{align}\label{eq:seesawProcesses}
  pp &\to AS,\ AS^{\pm},\ S^{\pm}S,\ S^{\pm} S^{\mp},\ S^{\pm\pm} S^{\mp},\ S^{\pm\pm} S^{\mp\mp};
\end{align}
with matrix elements including up to two additional partons. This is a large number of processes, and we cannot use generator biases because of the mix of final states, so we generate $2 \times 10^7$ events for each of 215 parameter points for all analyses. We show the total $m_{\ell\ell}$ distribution found in an event sample with $m_{S^{++}} = 170$ GeV, $m_{S^{\pm}} = 162$ GeV in Figure~\ref{fig:typeIImll}. Since we select only pairs of leptons with opposite signs, the topologies are similar to those of chargino pair production, but with BSM scalars instead of fermions. The $m_{\ell\ell}$ distribution is thus similar to that case, with a peak around $\Delta m = m_{S^{\pm}} - m_S$. In the end, we find total cross sections of $\mathcal{O}(1)$~pb, yet the efficiencies for any given signal region are barely more than $10^{-6}$ for the soft-lepton analysis: they are substantially smaller than the equivalents found for chargino pair production. This is presumably because of the isolation requirements: the processes with $S^{\pm\pm}$ (which have the largest cross sections) contain two off-shell $W$ decays, one of which should be hadronic (so as not to produce same-sign lepton pairs).

The result of these poor efficiencies is a failure of all analyses to constrain the type-II seesaw model in this parameter space. This result is noteworthy by itself since (as mentioned above) this space was previously identified as unconstrained by the LHC. But this situation also makes an exclusion plot impracticable. In lieu of a plot, we provide in Table~\ref{tab:typeIIsigPointInfo} the results for one point in our scan to which all four analyses\footnote{Recall once more that a limit can only be computed based on the most sensitive individual signal region for the ATLAS monojet analysis.} show at least some sensitivity. This point has $(m_{S^{\pm}},\Delta m(S^{\pm},S)) = (95,5)$~GeV and thus corresponds to $\lambda_4 = -0.061$. Table~\ref{tab:typeIIsigPointInfo} shows that, as for all points in our scan, there is neither an expected nor an observed limit; \ie, all $\text{CL}_s > 0.05$. Nevertheless, the ATLAS $2\ell$ and CMS monojet analyses still exhibit excesses visible in the form of expected $\text{CL}_s$ values significantly smaller than their observed counterparts. On the other hand, surprisingly, the ATLAS monojet search seems to show the opposite behavior by this metric, though insufficient statistics may be at play here. Meanwhile, while the significance is computed in the usual way for each analysis, we see nothing significant other than $p = 0.050$ for the ATLAS $2\ell$ analysis, which is however associated with a signal strength of $\hat{\mu} = 2.06$. Thus even in the best case we can find, this model produces insufficient events passing these analysis selections. By any measure, the type-II seesaw model does not fit the compressed LHC excesses.

\renewcommand{\arraystretch}{1.1}
\begin{table}
\centering
    \begin{tabular}{l l@{\hspace{3ex}}l@{\hspace{3ex}}l@{\hspace{3ex}}c}
& $\text{CL}_s^{\text{exp}}$ & $\text{CL}_s^{\text{obs}}$ & $p$-value & $\hat{\mu}$ \\
\midrule
ATLAS $2\ell$ & 0.387 & 0.620 & 0.050 & 2.06 \\
ATLAS $3\ell$ & 0.611 & 0.615 & $\geq 0.5$ & 0.62 \\
CMS monojet & 0.270 & 0.395 & 0.081 & 1.17 \\
ATLAS monojet & 0.553 & 0.399 & 0.277 & 0.72 
\end{tabular}
\caption{Statistics for the point $(m_{S^{\pm}},\Delta m(S^{\pm},S)) = (95,5)$~GeV in the type-II seesaw model. $\text{CL}_s \leq 0.05$ indicates exclusion at 95\%~CL.}\label{tab:typeIIsigPointInfo}
\end{table}
\renewcommand{\arraystretch}{1.0}

%% file: Sections/6_Conclusion.tex
\section{Conclusions}
\label{s6}

We have investigated the compatibility of several supersymmetric and non-supersymmetric models with the soft-lepton and monojet excesses at the LHC, and attempted to connect those excesses with the dark matter relic abundance. The competing models try to explain the excesses through processes with differing topologies, which lead to different distributions of the invariant mass of dileptons, and also to final states with differing average numbers of jets. Taken together, of the cases that we considered, we find that models producing lepton pairs via the decay of a new-physics state through an off-shell $Z$ boson exhibit the best compatibility. The other models that we considered feature decays of new particles via off-shell $W$ bosons or directly to leptons through a two-body process, and both of these cases produce wider distributions of $m_{\ell\ell}$ and possibly events with more jets. This therefore leads to poorer fits of the excesses, and even to cases where the observed limits are stronger than the expected ones. In particular, our results tend to favor the bino-wino MSSM and singlino-higgsino NMSSM explanations for the soft-lepton excesses. But the NMSSM produces not nearly enough monojet events, and we have found that both the doublet vector-like lepton and type-II seesaw models yield even worse fits. This altogether leaves the MSSM as the best option currently on the table. But it too leaves something to be desired.

While we have studied an array of well motivated and (crucially) distinct models, this work is by no means comprehensive. There exist multiple well known BSM scenarios worthy of future examination, including the various $N$-Higgs-doublet models. Alternatively, some future work on this subject should be aimed at constructing new models optimized from the bottom up to fit these excesses. It would be interesting to consider non-supersymmetric models where decays occur via an off-shell $Z$ boson, or via an off-shell BSM scalar: we are particularly motivated to examine new ways to produce relatively sharply peaked $m_{\ell\ell}$ distributions that are not identical to those seen in this work. We therefore hope to return to these ideas in the near future.

The adaptation of our recasts to \madanalysis, and a recast of the CMS soft-lepton search \cite{CMS:2021edw}, are currently ongoing. The latter will allow a comparison of the soft-lepton excesses between the two collaborations, and in particular will help us answer whether the apparent qualitative overlap between ATLAS and CMS soft-lepton excesses is accompanied by compatible signal model fits. At that point, it will also be worthwhile to perform a full statistical combination\footnote{We have not done that here for three reasons. (1) We did not want to obscure the fits for the different analyses, especially for points that did not fit well. (2) Without the CMS soft-lepton analysis, we would undersell the total significance. (3) Even using a powerful GPU on a machine with more than 100~GB of RAM, the statistical analysis using \pyhf takes $\mathcal{O}(5)$ minutes per point; this will only be multiplied in complexity for the combination, and for technical reasons that machine could not run \spey due to the too recent {\tt python} version running on it.} of the ATLAS soft-lepton, CMS soft-lepton and CMS monojet searches using a joint likelihood (since the analyses and signal regions are all independent; but sadly excluding the ATLAS monojet search) in order to obtain the overall best-fit points to all the different excesses and provide potentially provocative discovery $p$-values for excluding the Standard Model.

%% file: Sections/A_ATLAS_3l_extra_tables.tex
\section{Selections and yields for the inclusive off-shell \emph{WZ} selection}
\label{a1}

This appendix (starting on the next page) contains the definitions of the inclusive signal regions for the $3\ell$ analysis, ATLAS-SUSY-2019-09, and the yields and discovery $p$-values for each inclusive signal region.

\renewcommand{\arraystretch}{1.1}
\begin{table*}
\begin{center}\renewcommand{\arraystretch}{1.4}\setlength{\tabcolsep}{8pt}
\begin{tabular}{l c c c c c c}
\rule{0pt}{\dimexpr.7\normalbaselineskip+1mm}
& \multicolumn{4}{c}{\incSRhighnj} & \\[0.1cm]
& \texttt{a}      & \texttt{b}       & \texttt{c1}     & \texttt{c2} & \\\cline{2-5}
\mllmin [GeV]\ \ \ \ \ \ & $[1,12]$ & $[12,15]$ & $[1,20]$ & $[15,20]$ & \\
\rule{0pt}{\dimexpr.7\normalbaselineskip+1mm}
& \SRhighnjd{a} & \SRhighnjd{b} & \SRhighnjd{a-c} & \SRhighnjd{c} & \\[0.1cm]\bottomrule
\rule{0pt}{\dimexpr.7\normalbaselineskip+1mm}
& \multicolumn{2}{c}{\incSRlow} & \multicolumn{2}{c}{\incSRhigh} & \\[0.1cm]
\rule{0pt}{\dimexpr.7\normalbaselineskip+1mm}
& \texttt{b}       & \texttt{c}                  & \texttt{b}       & \texttt{c}       &   \\\cline{2-5}
\mllmin [GeV] & $[12,15]$ & $[12,20]$            & $[12,15]$ & $[12,20]$ & \\
\rule{0pt}{\dimexpr.7\normalbaselineskip+1mm}
& \makecell{\SRlowzjd{b},\\ \SRlownjd{b}} & \makecell{\SRlowzjd{b-c},\\ \SRlownjd{b-c}} & \makecell{\SRhighzjd{b},\\ \SRhighnjd{b}} & \makecell{\SRhighzjd{b-c},\\ \SRhighnjd{b-c}} \\[0.1cm]\bottomrule
& \multicolumn{5}{c}{\incSRlowhigh} \\
& \texttt{d} & \texttt{e1} & \texttt{e2} & \texttt{f1} & \texttt{f2} \\\cline{2-6}
\mllmin [GeV] & $[12,30]$ & $[12,40]$ & $[20,40]$ & $[12,60]$ & $[30,60]$ \\
\rule{0pt}{\dimexpr.7\normalbaselineskip+1mm}
& \makecell{\SRlowzjd{b-d},\\ \SRlownjd{b-d},\\ \SRhighzjd{b-d},\\ \SRhighnjd{b-d}} & \makecell{\SRlowzjd{b-e},\\ \SRlownjd{b-e},\\ \SRhighzjd{b-e},\\ \SRhighnjd{b-e}} & \makecell{\SRlowzjd{c-e},\\ \SRlownjd{c-e},\\ \SRhighzjd{c-e},\\ \SRhighnjd{c-e}} & \makecell{\SRlowzjd{c-f2},\\ \SRlownjd{c-f2},\\ \SRhighzjd{c-f2},\\ \SRhighnjd{c-f}} &  \makecell{\SRlowzjd{e-f2},\\ \SRlownjd{e-f2},\\ \SRhighzjd{e-f2},\\ \SRhighnjd{e-f}} \\[0.1cm] \bottomrule
& \multicolumn{4}{c}{\incSRlowhigh} & \\
& \texttt{g1} & \texttt{g2} & \texttt{g3} & \texttt{g4} & \\\cline{2-5}
\mllmin [GeV] & $[12,75]$ & $[30,75]$ & $[40,75]$ & $[60,75]$ & \\
\rule{0pt}{\dimexpr.7\normalbaselineskip+1mm}
& \makecell{\SRlowzjd{b-g2},\\ \SRlownjd{b-g2},\\ \SRhighzjd{b-g2},\\ \SRhighnjd{b-g}} & \makecell{\SRlowzjd{e-g2},\\ \SRlownjd{e-g2},\\ \SRhighzjd{e-g2},\\ \SRhighnjd{e-g}} & \makecell{\SRlowzjd{f1-g2},\\ \SRlownjd{f1-g2},\\ \SRhighzjd{f1-g2},\\ \SRhighnjd{f1-g}} & \makecell{\SRlowzjd{g1-g2},\\ \SRlownjd{g1-g2},\\ \SRhighzjd{g1-g2},\\ \SRhighnjd{g}} & \\
\end{tabular}
\end{center}
\caption{Summary of the selection criteria defining the inclusive SRs in the \ofs \WZ selection.}
\label{tab:ofs:disc}
\end{table*}
\renewcommand{\arraystretch}{1.0}

\begin{table*}
\centering
\renewcommand{\arraystretch}{1.1}
\begin{tabular}{l l r@{~$\pm$~\hspace{0ex}}l l}
SR & $N_{\mathrm{obs}}$ & \multicolumn{2}{c}{$N_{\mathrm{exp}}$} & $p(s=0)$ \\
\midrule
\incSRhighji{nja}  & $3$   &  $6.0$ & $1.6$\ \ \ \ \ & $ 0.50$ \\[0.04cm]
\incSRhighji{njb}  & $2$   &  $1.4$ & $0.6$ & $ 0.30$ \\[0.04cm]
\incSRhighji{njc1} & $7$   &  $9.5$ & $2.2$ & $ 0.50$ \\[0.04cm]
\incSRhighji{njc2}\ \ \ \ \ & $2$   &  $2.1$ & $0.8$ & $ 0.50$ \\[0.04cm]
\incSRlowji{b}     & $31$  &  $36$  & $4$   & $ 0.50$ \\[0.04cm]
\incSRhighji{b}    & $3$   &  $3.0$ & $0.9$ & $ 0.50$ \\[0.04cm]
\incSRlowji{c}     & $86$  &  $88$  & $7$   & $ 0.50$ \\[0.04cm]
\incSRhighji{c}    & $9$   &  $9.3$ & $1.5$ & $ 0.50$ \\[0.04cm]
\incSRj{d}         & $202$ &  $184$ & $12$  & $ 0.16$ \\[0.04cm]
\incSRj{e1}        & $332$ &  $308$ & $17$  & $ 0.16$ \\[0.04cm]
\incSRj{e2}        & $298$ &  $269$ & $15$  & $ 0.10$ \\[0.04cm]
\incSRj{f1}        & $479$ &  $457$ & $22$  & $ 0.23$ \\[0.04cm]
\incSRj{f2}        & $277$ &  $272$ & $13$  & $ 0.37$ \\[0.04cm]
\incSRj{g1}        & $620$ &  $593$ & $28$  & $ 0.21$ \\[0.04cm]
\incSRj{g2}        & $418$ &  $408$ & $20$  & $ 0.32$ \\[0.04cm]
\incSRj{g3}        & $288$ &  $285$ & $16$  & $ 0.38$ \\[0.04cm]
\incSRj{g4}        & $141$ &  $136$ & $10$  & $ 0.35$ \\[0.04cm]
\end{tabular}
\caption{Observed ($N_{\text{obs}}$) yields after the discovery fit and expected ($N_{\text{exp}}$) after the background-only fit, for the inclusive SRs of the \ofs \WZ selection. The last column indicates the discovery $p$-value ($p(s = 0)$). If the observed yield is below the expected yield, the $p$-value is capped at 0.5.}
\label{tab:results:offShell_discSRs}
\end{table*}